\documentclass[letterpaper, 10 pt, conference]{ieeeconf}  

\IEEEoverridecommandlockouts                              

\usepackage{graphicx}

\usepackage{amssymb}
\usepackage{pifont}
\usepackage{amsmath}
\usepackage{multirow}
\usepackage{array}

\newcommand{\finalcells}[2]{%
	\begingroup\sbox0{\begin{minipage}{3cm}\raggedright#1\end{minipage}}%
	\sbox2{\begin{minipage}{3cm}\raggedright#2\end{minipage}}%
	\xdef\finalheight{\the\dimexpr\ht0+\dp0+\smallskipamount\relax}%
	\xdef\finalheightB{\the\dimexpr\ht2+\dp2+\smallskipamount\relax}%
	\ifdim\finalheightB>\finalheight
	\global\let\finalheight\finalheightB
	\fi\endgroup
	\begin{minipage}[t][\finalheight][t]{3cm}\raggedright#1\end{minipage}&
	\begin{minipage}[t][\finalheight][t]{3cm}\raggedright#2\end{minipage}}

\title{\LARGE \bf
SoftEnNet: Symbiotic Monocular Depth Estimation and Lumen Segmentation for Colonoscopy Endorobots
}

\author{Alwyn Mathew$^{1}$, Ludovic Magerand$^{2}$, Emanuele Trucco$^{2}$ and Luigi Manfredi$^{1*}$
\thanks{This work was supported by the UK Engineering and Physical Sciences Research Council (EPSRC) grant number EP/W00433X/1}
\thanks{$^{1}$Alwyn Mathew and Luigi Manfredi are with the Division of Imaging Science and Technology, School of Medicine, University of Dundee, UK.
	$^{*}${\tt\small l.manfredi@dundee.ac.uk}
    }%
\thanks{$^{2}$Ludovic Magerand and Emanuele Trucco are with the Discipline of Computing, School of Science and Engineering, University of Dundee, UK.
    }%
}

\begin{document}

\maketitle
\thispagestyle{empty}
\pagestyle{empty}

\begin{abstract}
	
	Colorectal cancer is the third most common cause of cancer death worldwide. Optical colonoscopy is the gold standard for detecting colorectal cancer; however, about 25 percent of polyps are missed during the procedure. A vision-based autonomous endorobot can improve colonoscopy procedures significantly through systematic, complete screening of the colonic mucosa. The reliable robot navigation needed requires a three-dimensional understanding of the environment and lumen tracking to support autonomous tasks. We propose a novel multi-task model that simultaneously predicts dense depth and lumen segmentation with an ensemble of deep networks. The depth estimation sub-network is trained in a self-supervised fashion guided by view synthesis; the lumen segmentation sub-network is supervised. The two sub-networks are interconnected with pathways that enabling information exchange and thereby mutual learning. As the lumen is in the image's deepest visual space, lumen segmentation helps with the depth estimation at the farthest location. In turn, the estimated depth guides the lumen segmentation network as the lumen location defines the farthest scene location. Unlike other environments, view synthesis often fails in the colon because of the deformable wall, textureless surface, specularities, and wide field of view image distortions, all challenges that our pipeline addresses. We conducted qualitative analysis on a synthetic dataset and quantitative analysis on a colon training model and real colonoscopy videos. The experiments show that our model predicts accurate scale-invariant depth maps and lumen segmentation from colonoscopy images in near real-time. 
	
\end{abstract}

\section{INTRODUCTION}

Colorectal cancer was the third most reported cancer in 2020 and the second most causing death worldwide \cite{xi2021global} in both genders representing 10\% of the global cancer incidence and 9.4\% of all cancer-caused deaths. Based on projections of aging, population growth, and human development, the global number of new CRC cases is expected to reach 3.2 million by 2040 \cite{sung2021global}. The growing number of CRC cases and rising incidence among younger generations continue to impose a significant financial burden and public health challenge. CRC incidence has decreased or stabilized in a few countries with high human development, owing primarily to a healthier lifestyle and the implementation of a screening program a decade ago. Early detection is a critical factor in preventing metastasis, lowering mortality, and improving future prognosis and quality of life. These statistical analyses emphasize the importance of a healthy lifestyle to reduce risk factors, CRC screening programs for early detection, and new research on CRC prevention and treatment. 

The gold standard procedure for CRC screening is optical colonoscopy because of its dual capability to visually inspect the colonic mucosa and remove polyps that may eventually become cancerous. This procedure is complex and causes pain and discomfort. The outcome is related to the skill of the operator \cite{baxter2009association}. Adenoma Detection Rate (ADR) is the proportion of patients with at least one colorectal adenoma detected among all patients examined by an endoscopist. The ADR is one of the key parameters that assess the outcome quality related to technical skills in performing the procedure. Autonomous endorobots can improve the quality and reduce the dependency of outcome on skills. Such robots are medical devices \cite{manfredi2021endorobots} that can conduct a systematic screening of the colonic mucosa with some level of autonomy. The critical tasks of an on-board vision system for navigation are identifying the advancing direction as well as building a 3D reconstruction of the colon wall to perform surgical tasks. Accuracy and fast processing are vital aspects for real-time control of an autonomous endorobot \cite{manfredi2019soft}.

\begin{figure*}
	\centering
	\includegraphics[width=0.7\linewidth]{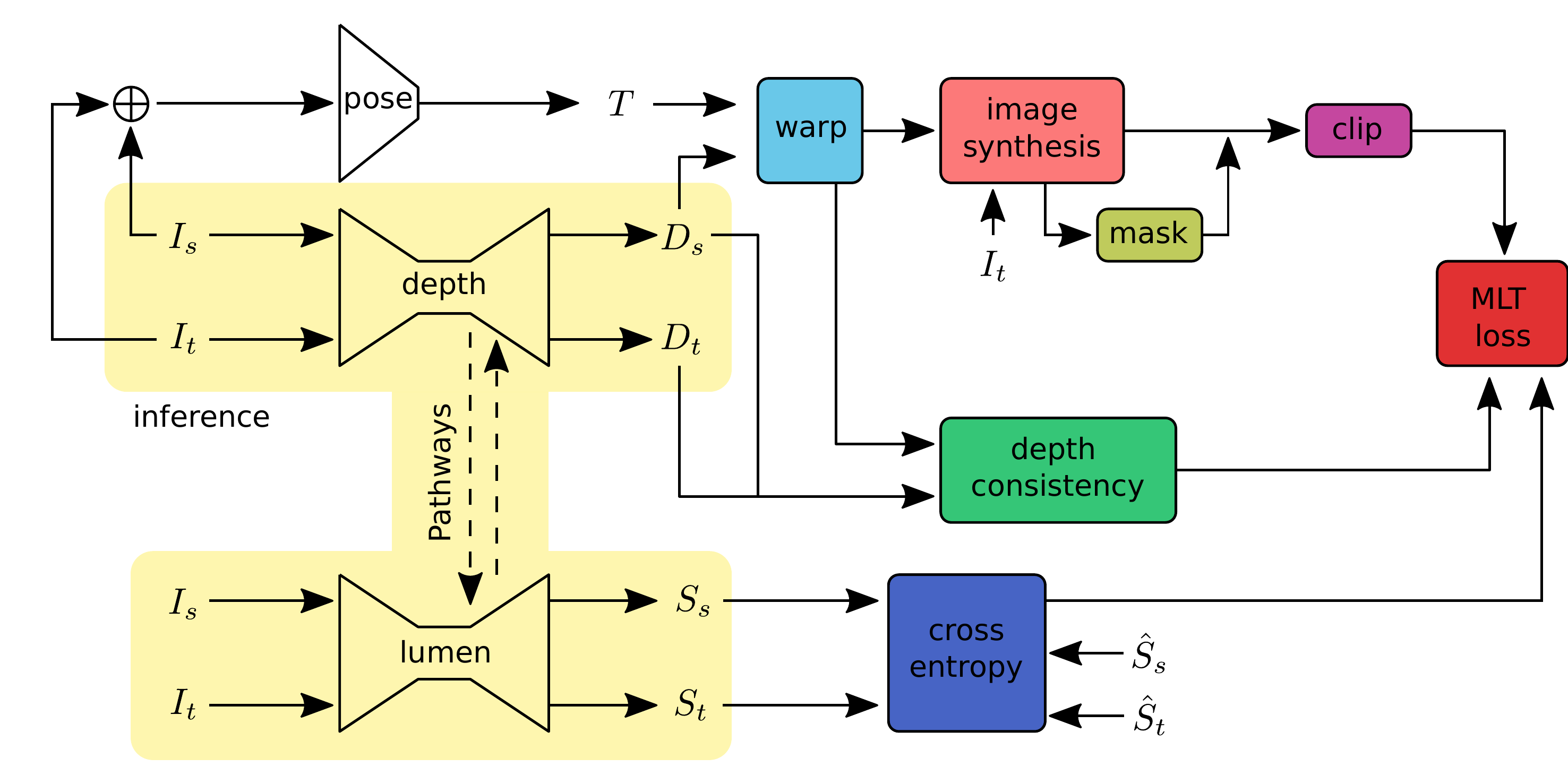} 
	\caption{\textbf{Multi-task training pipeline}: Input images ($I_s$, $I_t$) are successive frames from a video sequence. The \emph{depth} sub-network predicts independent per-pixel dense depth maps ($D_s$, $D_t$) form the source $I_s$ and target $I_t$ images.  The \emph{pose} sub-network takes the concatenated image sequence ($I_s$, $I_t$) as input and predicts the rotation $R$ and translation $t$ that defines the transformation $T$ between source and target images. The \emph{warp} function takes the predicted depths and poses to warp the source image coordinates to the target or vice versa. The \emph{image synthesis} projects the source image $I_s$ to the target image $I_t$ with the warped image coordinates. The synthesized images are compared with real images to construct the loss. The image synthesis loss outliers are removed with the \emph{clip} function. The \emph{lumen} sub-network locates the lumen ($S_s, S_t$) in the source $I_s$ and target $I_t$ images. The \emph{pathways} share relevant features between task networks. A cross-entropy loss is used to guide the lumen decoder head with ground truth lumen segmentation ($\hat{S}_s, \hat{S}_t$). The \emph{MLT} weights the task losses enforces homoscedastic task uncertainty. $\bigoplus$ denotes image concatenation. At inference (yellow), only depth and lumen sub-network will be used.}
	\label{fig:pwdnet_pipeline}
\end{figure*}

This work reports a real-time monocular depth estimation and lumen segmentation system for colonoscopy that can pave the way for implementing autonomous locomotion and surgical tasks. A monocular camera meets the constraint if limited space inside the colon that makes the use of a 3D stereo vision difficult. Advantage of the proposed study compared to previously reported work includes: 
\begin{enumerate}
	\item A novel multi-task network, SoftEnNet, that simultaneously predicting dense depth estimation and lumen segmentation for endo robot navigation.
	\item A symbiotic architecture between depth and lumen sub-networks in the multi-task SoftEnNet network that benefits both tasks.
	\item A novel validity mask that removes pixels not contributing to view synthesis from the loss calculation, those are affected by specular lighting, low texture surface, or violating the Lambertian assumption.
\end{enumerate}

\section{RELATED WORK}

\subsection{Depth Estimation}

Monocular dense depth estimation from videos has remarkably improved over the last few years, but the same doesn't apply to colonoscopy. Per-pixel depth estimation is still an open problem in colonoscopy mainly because of the low-textured, deformable and reflective colon surface. Supervised learning is not achievable as the colonoscope cannot carry additional depth sensors to gather ground truth depth maps due to the size restriction of the colon. The deformable nature of the colon makes data obtained from computer tomography (CT) of unreliable. Colon modeling clues and characteristic geometry has been previously explored for the 3D reconstruction of the colon. Hong \textit{et al.} \cite{hong20143d} used colon fold contours to estimate depth. The colon surface was reconstructed with a combination of structure-from-motion (SfM) and shape-from-shading (SfS) techniques in Zhao \textit{et al.} \cite{zhao2016endoscopogram} but resulted in inconsistent depth estimation. Images from chromoendoscopy that produced enhanced colon surface texture were used to improve SfM \cite{widya2020stomach, widya2019whole, widya2021self}. As chromoendoscopy is not very common and computationally expensive, these techniques have minimal application. Ma \textit{et al.} \cite{ma2019real} used an SfM approach to generate pseudo ground truth depth data, and that was used to train a deep learning network to predict depth. Ma \textit{et al.} \cite{freedman2020detecting} proposed a self-supervised depth estimation method from video and enforced temporal consistence between the frames. Synthetic data generated from photo-realistic phantoms and CT scans have been studied extensively to aid data-driven approaches \cite{mahmood2018deep, visentini2017deep}. These approaches are trained with rendered images, not in real colonoscopy images, and thus perform poorly on real data.

\subsection{Semantic Segmentation}

Most semantic segmentation of the colon target polyps, whereas our segmentation sub-network focuses on lumen segmentation which is essential for endorobot navigation. Most initial attempts have used traditional image processing techniques like watershed \cite{bernal2012towards}, Hough transform \cite{ganz2012automatic} and canny edge detection \cite{tajbakhsh2015automated}. Wang \textit{et al.} \cite{wang2018development} introduced real-time polyp segmentation with SegNet \cite{guo2019giana} and with fully convolutional dilation networks. Jha \textit{et al.} \cite{jha2019resunet++} introduced ResUNet++ that improved segmentation over UNet architecture. Wang \textit{et al.} \cite{wang2022boundary} proposed a boundary-aware neural network and Jha \textit{et al.} \cite{jha2020doubleu} introduced Double UNet architecture for segmentation. Huang \textit{et al.} \cite{huang2021hardnet} uses a low memory latency HarDNet backbone \cite{chao2019hardnet}  with a Cascaded Partial Decoder for fast and accurate polyp segmentation. Tomar \textit{et al.} \cite{tomar2022fanet} uses hard attention based on an iterative refinement method using Otsu thresholding. 

\begin{figure*}[thpb]
	\centering
	\includegraphics[width=0.7\linewidth]{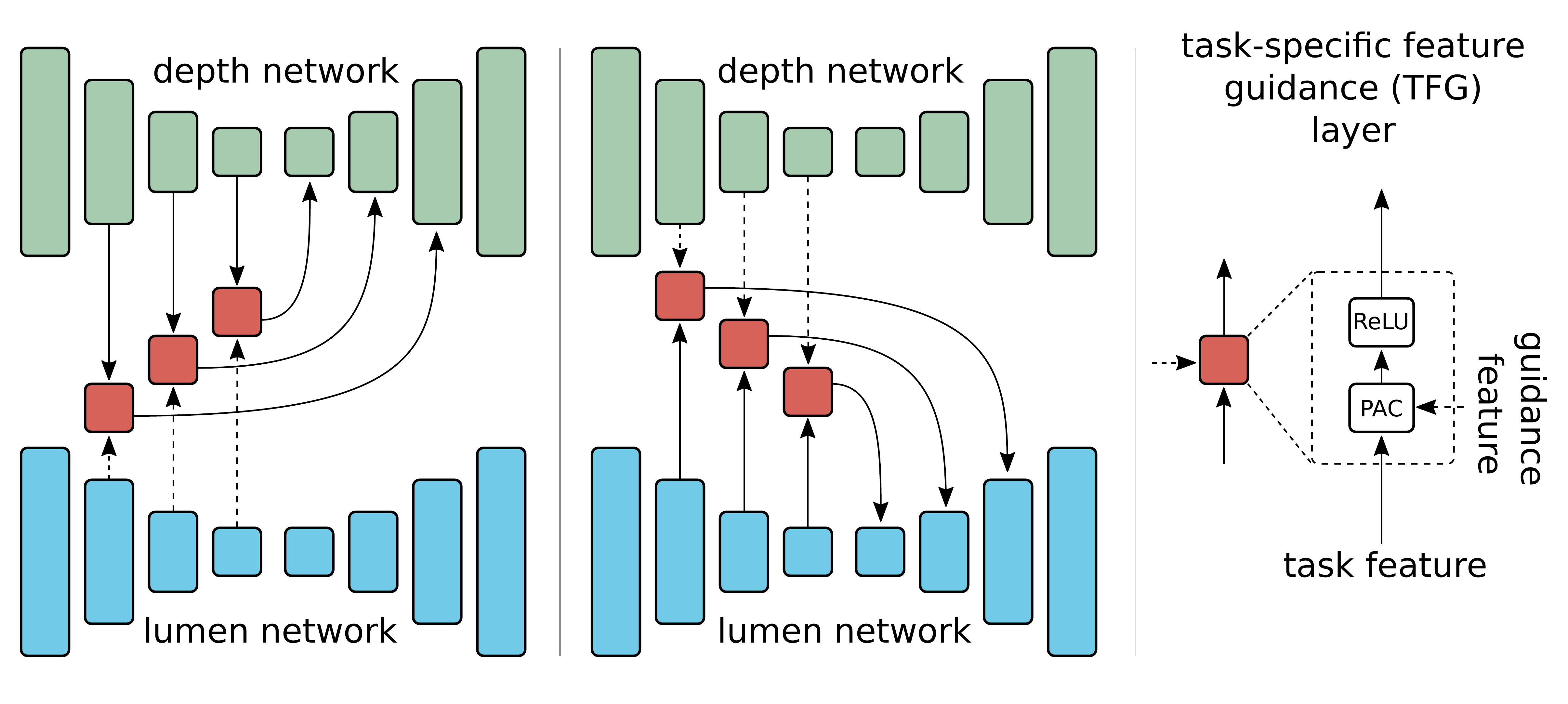}
	\caption{\textbf{SoftEnNet}: The multi-task network predicts depth and lumen location in the image simultaneously. SoftEnNet consists of two sub-networks for each task. These sub-networks are connected with \emph{task-specific feature guidance (TFG) layers} to create multiple pathways between networks sharing functional feature space. The guidance layers consist of a pixel-adaptive convolutional layer (\emph{PAC}) and Rectified Linear Unit (\emph{ReLU}). The guidance features are used to adapt \emph{PAC} kernels applied to the task features. Left: Pathways from lumen network to depth network. Center: Pathways from depth network to lumen network. Right: Task-specific feature guidance (TFG) layer.}
	\label{fig:pwdnet}
\end{figure*}

\section{METHODS}

The presented work simultaneously estimates the depth and segment the lumen in a colonoscopy RGB image stream with a novel multi-task network  (Fig.~\ref{fig:pwdnet_pipeline}). This network consists of three subnetworks: \textit{depth}, \textit{pose}, and \textit{lumen}. The self-supervised depth sub-network is guided by view synthesis loss between image frames. It takes a monocular image as input and outputs a dense depth map. The pose subnetwork predicts the rigid transformation $T_{t \rightarrow s}$ between two input image frames. The predicted depth $D_t$ along with transformation $T_{t \rightarrow s}$ can be used to synthesis target image $\hat{I}_t$ from source image $I_s$ with geometric projection. The lumen sub-network predicts the lumen segmentation in a supervised fashion. The lumen sub-network takes a single image and segments the lumen in the image. Pathways connecting the depth and lumen subnetworks enable mutual learning and combine existing skills to learn new tasks faster and more effectively. For inference, only the depth and lumen sub-networks are used.

\subsection{Self-Supervised Depth Estimation}

Our self-supervised monocular depth estimation framework is used in the proposed depth estimation framework with two networks:

\begin{itemize}
	\item Depth Network: $N_D: I_t \rightarrow D_t$ takes a monocular image $I_t$ and predicts corresponding dense depth maps $D_t$.
	\item Pose Network: $N_P: (I_t, I_s) \rightarrow T_{t \rightarrow s}$ takes two successive image frames ($I_t, I_s$) to predict the six degrees of freedom (DOFs) rigid transformation $T_{t \rightarrow s}$ between source $I_s$ and target $I_t$ image.
\end{itemize}

Self-supervised depth is achieved by using view synthesis \cite{garg2016unsupervised} between video frames as supervision. Below we explain the image synthesis function and its inverse ($\pi, \pi^{-1}$), image reconstruction loss $L_v$, depth consistency loss $L_{c}$, edge-guided smoothness loss $L_e$, and valid mask $V$ used in self-supervised depth and pose loss.

\subsubsection{Image synthesis} \label{synthesis}

Projection function $\pi$ maps the 3D point cloud to 2D image coordinates and unprojection function $\pi^{-1}$ maps the 2D image to 3D world coordinates. The target image $\hat{I_t}$ can be synthesized from source images $I_s$ by projecting image coordinates from target to source with rigid transformation $T_{t \rightarrow s}$, projection $\pi$ and unprojection $\pi^{-1}$ function, then inverse sampling pixel values with remap function $\zeta$. A point cloud $P_t$ can be generated from estimated depth $D_t$ for corresponding input image $I_t$ as follows:

\begin{equation}
\begin{gathered}
p_s = \pi(T_{t \rightarrow s} P_t), \quad P_t = \pi^{-1}(p_t, D_t), \quad \hat{I}_t^{ij} = \zeta(I_s^{i'j'})
\end{gathered}
\end{equation}

where $p_t$ is the set of image coordinate and $P_t$ is the 3D point cloud corresponding to target image $I_t$. The relative pose $T_{t \rightarrow s}$ from target to source image is estimated by the pose network. This relative pose can be used to transform target cloud point to source cloud point $P_s = T_{t \rightarrow s} P_t$. With the projection model $\pi$, the 3D cloud point $P_s$ can be mapped to source image coordinate $p_s$ as follows. Target image can be synthesized by inverse warping source image coordinate and sampling \cite{godard2017unsupervised} with $\zeta$. The projected source image coordinate $p_s(i', j')$ will be used to sample pixel values with sampling function $\zeta$ to generate target image $\hat{I}_t$. The differentiable spatial transformer network \cite{jaderberg2015spatial} is used to synthesize images with bilinear interpolation. Interpolation is used to find the approximate pixel values in the source image as the inverse transformed grid may not fall on image grid points.

\subsubsection{Image reconstruction loss} \label{photometric}

The photometric loss $L_v$ between the reference frame and the reconstructed frame act as a self-supervised loss for the depth network. Following previous works \cite{zhou2017unsupervised, mahjourian2018unsupervised, yin2018geonet}, we use Structural Similarity (SSIM) \cite{wang2004image} along with L1 for the photometric loss with weight $\eta$. This takes advantage of the ability of SSIM denoted as $S(.)$ to handle complex illumination change along with robust to outliers L1 loss as follows: 

\begin{multline}
	L_v^s(I_t, \hat{I}_t) = \frac{1}{|V|} \sum_{p \in V}^{} \eta \frac{1-S(I_t(p), \hat{I}_t(p))}{2} \\ + (1-\eta) | I_t(p) - \hat{I}_t(p) |
\end{multline}

Here, $|V|$ is the number of valid pixels in the validity mask $V$. Masking out pixels  with $V$ that violates non-zero optical flow and Lambertian surface assumption avoid propagation of corrupted gradients into the networks. The validity mask is computed as follows:

\begin{equation}
V = M_{ego} \cdot (| I_t(p) - \hat{I}_t(p) | < | I_t(p) - I_s(p) |)
\end{equation}

$M_{ego}$ is the binary mask of valid pixels similarly to \cite{mahjourian2018unsupervised}. The pixels that appear to be static between source and target images caused by no ego motion, special lighting, lens artifacts and low texture are marked invalid in $V$. Parts of photometric loss that violate the assumptions generate a large loss, yielding a large gradient which potentially worsens the performance. Following \cite{zhou2018unsupervised}, we clip the loss with $\theta$ which is the $p^{th}$ percentile of $L_v$, to removes the outliers as $L_v(s_i) = min(s_i, \theta)$ where $s_i$ is $i^{th}$ cost in $L_v$.

\begin{table*}[h]
	\caption{Quantitative study on SoftEnNet on UCL 2019 synthetic data. In \textit{Simple Pathways} model,  PAC are replaced by CONV in TFG layers. Best values are bold and second best are underlined.}
	\label{tab:quant}
	\begin{center}
		\begin{tabular}{|c|c|c|c|c|c|c|c|c|c|}
			\hline
			\multirow{2}{*}{\textbf{Model}} & \multicolumn{7}{c|}{\textbf{Depth}} & \multicolumn{2}{c|}{\textbf{Semantic}} \\
			\cline{2-10} 
			& Abs Rel $\downarrow$ & Sq Rel $\downarrow$ & RMSE $\downarrow$ & RMSE log $\downarrow$ & $\delta<1.25 \uparrow$ & $\delta<1.25^2 \uparrow$ & $\delta<1.25^3 \uparrow$ & \shortstack{Lumen \\IoU} $\uparrow$ & \shortstack{Mean \\IoU} $\uparrow$ \\
			\hline
			SfMLearner \cite{zhou2017unsupervised} 	   & 0.451 & 0.711 & 1.768 & 0.537 & 0.366 & 0.618 & 0.785 & - & - \\
			\hline
			Monodepth1 \cite{godard2017unsupervised} & 0.444 & 0.752 & 1.755 & 0.531 & 0.371 & 0.625 & 0.790 & - & - \\
			\hline
			DDVO \cite{wang2018learning}        & 0.452 & 0.772 & 1.768 & 0.538 & 0.366 & 0.618 & 0.785 & - & - \\
			\hline
			Monodepth2 \cite{godard2019digging} & 0.446 & 0.756 & 1.757 & 0.532 & 0.360 & 0.624 & 0.789 & - & - \\
			\hline
			HR Depth \cite{lyu2021hr} & 0.448 & 0.762 & 1.762 & 0.534 & 0.369 & 0.621 & 0.788 & - & - \\
			\hline
			AF-SfMLearner \cite{shao2021self} & 0.387 & 0.618 & 1.648 & 0.480 & 0.420 & 0.686 & 0.831 & - & - \\
			\hline
			AF-SfMLearner2 \cite{shao2022self} & 0.352 & 0.545 & 1.581 & 0.448 & 0.445 & 0.712 & 0.852 & - & - \\
			\hline
			Baseline Depth     & 0.332 & 0.565 & 1.364 & 0.376 & 0.497 & 0.789 & 0.913 & - & - \\
			\hline
			Baseline Lumen	   & -     & -     & -     & -     & -     & -     & -     & \underline{0.912} & \underline{0.954} \\
			\hline
			Simple Pathways    & \underline{0.150} & \underline{0.144} & \underline{0.720} & \underline{0.187} & \underline{0.814} & \underline{0.947} & \underline{0.981} & \textbf{0.931} & \textbf{0.964} \\  
			\hline
			SoftEnNet  & \textbf{0.145} & \textbf{0.112} & \textbf{0.650} & \textbf{0.184} & \textbf{0.823} & \textbf{0.954} & \textbf{0.983} & \textbf{0.931} & \textbf{0.964} \\  
			\hline
		\end{tabular}
	\end{center}
\end{table*}

\subsubsection{Depth consistency loss} \label{consist}

Depth consistency \cite{bian2021unsupervised} loss $L_c$ ensure that the depth maps $D_s, D_t$ estimated by the depth network with source and target images $I_s, I_t$ represent the same 3D scene. Minimizing this loss not only encourages depth consistency in the sequence but also in the entire sequence with overlapping image batches. Depth consistency loss is defined as:

\begin{equation}
\begin{gathered}
L_c^s(D_s, D_t) = \frac{1}{|V|} \sum_{p \in V}^{} DC^s \\
DC^s = \frac{ | \hat{D}_s^t(p) - \hat{D}_s(p) | }{\hat{D}_s^t(p) + \hat{D}_s(p)}
\end{gathered}
\end{equation}

where $\hat{D}_s^t$ is the projected depth of $D_t$ with pose $T_{t \rightarrow s}$ which corresponds to depth at image $I_s$ and $\hat{D}_s^t$ need to be compared with $D_s$ but cannot be done as the projection does not lie in the grid of $D_s$. Thus, $\hat{D}_s$ is generated using the differentiable bilinear interpolation on $D_s$ and compared with $\hat{D}_s^t$. The inverse of $DC$ will act as occlusion mask $M_o$ that can mask out occluded pixels from source to target view, mainly around the haustral folds of the colon. The image reconstruction loss $L_s$ is re-weighted with $M_o$ mask as follow:

\begin{equation}
\begin{gathered}
{L^s_v}' = \frac{1}{|V|} \sum_{p \in V}^{} M_o^s \cdot L^s_v \\
M_o^s = 1 - DC^s 
\end{gathered}
\end{equation}

\subsubsection{Edge-guided smoothness loss} \label{smoothness}

The depth maps are regularized with an edge guided smoothness loss as homogeneous and low texture region does not induce information in image reconstruction loss. The smoothness loss is adapted from \cite{wang2018learning} as follows:

\begin{equation}
L_e^s(I_t, D_t) = \sum_{p \in I}^{} e^{-\triangledown I_t(p)} \triangledown D'_t(p)
\end{equation}

where $\triangledown$ denotes first order gradient and $D'_t= (1/D_t)/mean(1/D_t)$ means normalized inverse depth. To avoid the optimizer getting trapped in local minima, the depth estimation and loss are computed at multiple scale \cite{zhou2017unsupervised, godard2019digging}. The total loss is averaged over scales (S=4) and image batches:

\begin{equation}
L_d = \frac{1}{S}\sum_{s=1}^{S} \sigma_1 {L^s_v}' + \sigma_2 L_c^s + \sigma_3 L_e^s
\end{equation}

\subsection{Lumen Segmentation}

Lumen segmentation is defined as a per-pixel class assignment. Each pixel is assigned a class label $c \in C = \{0, 1, 2\}$. where 0 indicates colon wall, and 1 indicates lumen. The lumen segmentation network $N_L: I_t \rightarrow S_t$ is trained in a supervised manner with ground truth lumen maps $\hat{S}_t, \hat{S}_s$. This network predicts a posterior probability $S$ that a pixel belongs to a class $c \in C$, which is then compared to the one-hot encoded ground truth labels $\hat{S}$ inside the cross-entropy loss as follows:

\begin{equation}
L_l = - \sum_{c \in C}^{} \hat{S}_{t, c} \cdot log(S_{t, c}) + \hat{S}_{s, c} \cdot log(S_{s, c})
\end{equation}

\subsection{Multi-Task Joint Optimization}

Multi-task learning involves the problem of optimizing a model for multiple objectives. The naive approach to combining multi-objective losses would be to perform a weighted linear sum of the losses for each task:

\begin{equation}
L = w_d L_d + w_l L_l
\end{equation}

where $w_d$ and $w_l$ are weights corresponding to depth estimation and lumen segmentation task, respectively. However, performance would be susceptible to weight value $w$ and these hyper-parameters are expensive to tune. We adopt a more convenient approach to learning the optimal weights: we follow \cite{kendall2018multi} to weight our multi-task objectives by considering the homoscedastic uncertainty of each task:

\begin{equation}
L = \frac{1}{2\gamma_1^2} L_d + \frac{1}{2\gamma_2^2} L_l + log(1+\gamma_1) + log(1+\gamma_2)
\end{equation}

Homoscedastic uncertainty \cite{kendall2018multi} is task-specific and does not change with varying input. Increasing the noise parameter lowers the weight for the particular task with learnable parameters $\gamma$. The objective optimizes a more significant uncertainty, which should result in a minor contribution of the task's loss to the total loss. This lets us simultaneously learn quantities with different units or scales in depth and lumen estimation. The noise parameter $\gamma_1$ and $\gamma_2$ weights depth and lumen, respectively. During our experiments we found that $\gamma_1$ is quite low compared to $\gamma_2$.

\section{NETWORK ARCHITECTURE}

We propose a novel symbiotic multi-task SoftEnNet that simultaneously predicts depth, pose, and lumen segmentation. SoftEnNet consists of three subnetworks: Depth, Pose, and Lumen. As related tasks like depth estimation and lumen segmentation could benefit from each other, depth and lumen networks are connected with pathways to facilitate information exchange as shown in Fig. \ref{fig:pwdnet}. Depth at the farthest visual space in an image is defined as the lumen. Lumen prediction will improve with dense depth estimation, while depth at the farthest points will improve with lumen segmentation. Multi-scale pathway connections are built between sub-networks that enable the exchange of task-specific features at coarse to fine scales. 

We defined the task-specific feature guidance (TFG) layer in each pathway connection to inject depth information into the lumen network and lumen segmentation information into the depth network as shown in Fig. \ref{fig:pwdnet}. Inspired by \cite{guizilini2020semanticallyguided}, the TFG layer leverages the knowledge from the guidance task to the target task network using pixel-adaptive convolutions \cite{su2019pixel} (PAC). The PAC layer addresses some limitations inherent in standard convolution operations, such as translation invariance, which makes it content-agnostic. The TFG layers in depth and lumen pathways generate depth-aware semantic and semantic-aware depth feature representations. These guided feature maps disambiguate semantically, and depth features, resulting in better feature space for prediction. The TFG layer consists of a PAC layer followed by ReLU \cite{agarap2018deep}. To ensure a better starting condition for further training, we scale the guiding features by a constant PAC-$\alpha$. In the initial training epochs, the guidance feature is scaled by a small constant, making it behave very similarly to the traditional convolutional layer. Later during the training, PAC-$\alpha$ is increased gradually to 1.

The \textit{depth sub-network} is inspired by \cite{godard2019digging} with ResNet18 \cite{he2016deep} as the backbone and multiscale decoder that predicted depth at four different scales. The \textit{lumen sub-network} follows a similar U-Net architecture as the depth network, with ResNet18 as the backbone and decoder that predicted lumen segmentation at the highest scale. Following \cite{godard2019digging}, the \textit{pose sub-network} shares the depth encoder weights, and the decoder consists of a sequence of four convolutional layers.

\begin{table}[h]
	\caption{Ablation study on task weight on UCL 2019 synthetic data. \textit{Depth} and \textit{Lumen} indicate baseline depth and lumen segmentation networks. \textit{L} denotes learned task weights.}
	\label{tab:task_wt}
	\begin{center}
		\resizebox{\linewidth}{!}{
			\begin{tabular}{|c|c|c|c|c|c|}
				\hline
				\multirow{2}{*}{\textbf{Model}} & \multirow{2}{*}{$\boldsymbol{w_d, w_l}$} & \multicolumn{2}{c|}{\textbf{Depth}} & \multicolumn{2}{c|}{\textbf{Semantic}} \\
				\cline{3-6}
				& & RMSE $\downarrow$ & $\delta<1.25 \uparrow$ & \shortstack{Lumen \\IoU} $\uparrow$ & \shortstack{Mean \\IoU} $\uparrow$\\
				\hline
				Depth 		& -	  			& 0.681 	& 0.811 		 	& - 	 				& -		   \\
				\hline
				Lumen 	    & -	  			& -  		& - 				& 0.912 	 			& 0.954    \\
				\hline
				SoftEnNet	& 0.4, 0.6   	& 0.679 	& 0.815 			& 0.901 				& 0.948 \\
				\hline
				SoftEnNet	& 0.6, 0.4   	& 0.661 	& 0.814 			& 0.919 				& 0.958 \\
				\hline
				SoftEnNet	& 0.7, 0.3   	& 0.669 	& 0.819 			& 0.920 				& 0.958 \\
				\hline
				SoftEnNet	& 1.0, 1.0   	    & 0.681 	& 0.813 			& 0.917 				& 0.956 \\
				\hline
				SoftEnNet	& L     		& \textbf{0.650} 	& \textbf{0.823} & \textbf{0.931} 	& \textbf{0.964} \\
				\hline
		\end{tabular}}
	\end{center}
\end{table}

\begin{table}[h]
	\caption{TFG PAC-$\alpha$ Ablation study on UCL 2019 synthetic data. $0.0001 \rightarrow 1$ denotes that the PAC-$\alpha$ is increase from $0.0001$ to $1$ during training.}
	\label{tab:pac_alpha}
	\begin{center}
		\begin{tabular}{|c|c|c|c|c|}
			\hline
			\multirow{2}{*}{\textbf{PAC-}$\boldsymbol{\alpha}$} & \multicolumn{2}{c|}{\textbf{Depth}} & \multicolumn{2}{c|}{\textbf{Lumen}} \\
			\cline{2-5}
			&	RMSE $\downarrow$ & $\delta<1.25 \uparrow$ & \shortstack{Lumen \\IoU} $\uparrow$	& \shortstack{Mean \\IoU} $\uparrow$ \\
			\hline
			0.0001	  	& 0.685 		  & 0.816    		  & 0.935 					& 0.966\\
			\hline
			0.001	  	& 0.658 		  & 0.821    	 	  & 0.936 					& 0.966\\
			\hline
			0.01	  	& 0.646 		  & 0.820    		  & 0.936 					& 0.966\\
			\hline
			0.1	  		& 0.888 		  & 0.811    		  & 0.928 					& 0.962\\
			\hline
			1.0	  		& 0.761 		  & 0.788    		  & 0.928 					& 0.963\\
			\hline
			0.0001 $\rightarrow$ 1	& 0.650 		  & 0.823    	 	  & 0.931 					& 0.964\\
			\hline
		\end{tabular}
	\end{center}
\end{table}

\section{EXPERIMENTS}

\subsection{Implementation details}

The models are built using Pytorch \cite{paszke2017automatic} with Adam \cite{kingma2014adam} optimizer to minimize the training error. The model is trained on an Nvidia A6000 for 40 epochs with a batch size of 20.   The initial learning rate of the model is $0.0001$, which is then cut in half after 30 epochs. The sigmoid output $d$ of the depth network is scaled as $D=1 / (d \cdot 1/(d_{max} - d_{min}) + 1/d_{min})$, where $d_{min}=0$ and $d_{max}=20.0$ units correspond to $0-20cm$. UCL 2019 synthetic dataset \cite{rau2019implicit} image network input image size is $256 \times 256$ (height $\times$ width) and our CTM dataset is $512 \times 768$. These hyperparameters are empirically determined: $\eta = 0.85$, $\sigma_1=1.0$, $\sigma_2=0.5$, and $\sigma_3=0.001$. We perform horizontal flipping on 50\% of the images in the UCL 2019 dataset. Because of the off-centered principal point of the wide-angle camera, image flipping is not performed in the CTM dataset. The clip loss $p=95$ following \cite{zhou2018unsupervised}. PAC-$\alpha$ is initially set to $0.0001$ and doubled every 5 epochs until it reaches $1$.

\begin{figure*}
	\centering
	\includegraphics[width=\linewidth]{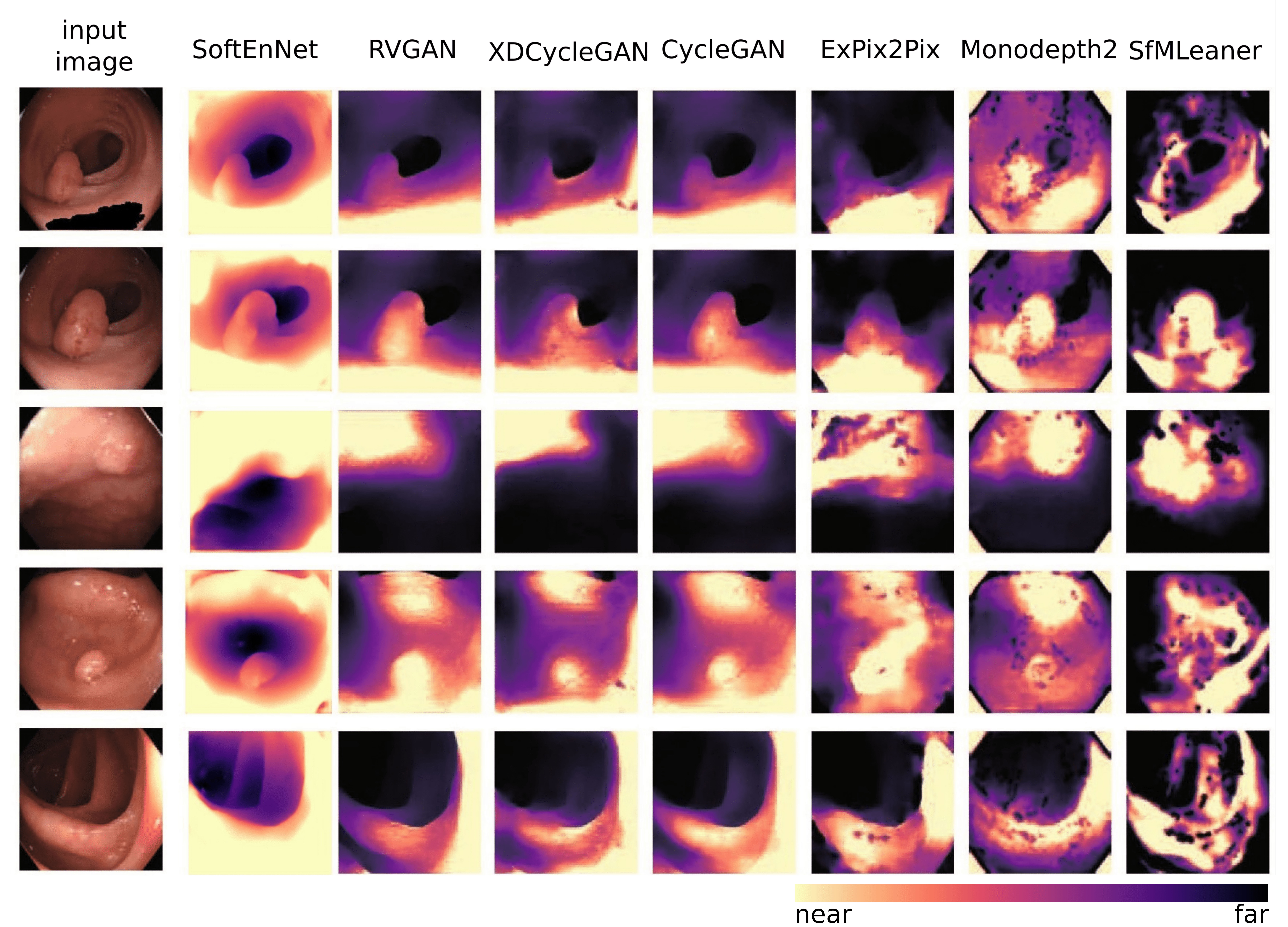} 
	\caption{Qualitative comparison of \textbf{SoftEnNet} with RVGAN \cite{itoh2021unsupervised}, XDCycleGAN \cite{mathew2020augmenting}, CycleGAN \cite{zhu2017unpaired}, ExPix2Pix \cite{rau2019implicit}, Monodepth2 \cite{godard2019digging} and SfMLearner \cite{zhou2017unsupervised} on RVGAN real colonoscopy images.}
	\label{fig:qual_2}
\end{figure*}

\subsubsection{UCL 2019 synthetic dataset}

The UCL synthetic data \cite{rau2019implicit} is created using human CT colonography and manual segmentation and meshing.
The Unity engine is used to generate RGB images of size $256 \times 256$ with a virtual pinhole camera ($90^\circ$ FOV, two attached light sources) and depth maps with a maximum range of 20cm. The dataset contains 15,776 RGB images, and ground truth depth maps divided into train, validation, and test sets with ratio 8:1:1. The dataset is divided into nine subsets, each with three lighting conditions and one of three materials. Light sources differ in spot angle, range, color, and intensity, while materials differ in color, reflectiveness, and smoothness. Lumen segmentation ground truth was extracted from ground truth depth maps by clipping the depth map at 95 percentile depth.

\subsubsection{Colonoscopy training model dataset}

The Colon Training Model (CTM) dataset was captured using an off-the-shelf full HD camera ($1920 \times 1080$, 30Hz, $132^\circ$ HFOV, $65^\circ$ VFOV, $158^\circ$ DFOV, FID 45-20-70, white ring light around camera) inside a plastic phantom used for medical professional training. The phantom simulates the 1:1 anatomy of the human colon, including internal diameter, overall length, and haustral folds, as seen in an optical colonoscopy. While recording videos, the camera is attached to a linear actuator that moves forward and backward with a constant speed of approximately 2 cm/s. The dataset contains approximately 10,000 images divided into an 8:2 train and validation set with varying internal and external light settings, but no ground truth depth maps are provided. Lumen segmentation pseudo ground truths are extracted from baseline depth predicted by clipping at 95 percentile depth.

\begin{figure}[]
	\centering
	\resizebox{\linewidth}{!}{
		\begin{tabular}{c c @{\hspace{0.1cm}} c @{\hspace{0.1cm}} c @{\hspace{0.1cm}} c}
			
			&
			\shortstack{Input \\image} & 
			\shortstack{Baseline \\depth} & 
			\shortstack{SoftEnNet \\depth} & 
			\shortstack{SoftEnNet \\ lumen} \\
			
			{\rotatebox{90}{\hspace{3.5mm}\shortstack{UCL\\ Syn\\ \cite{rau2019implicit}}}} &
			\includegraphics[width=0.2\linewidth]{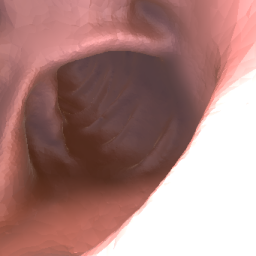} & \includegraphics[width=0.2\linewidth]{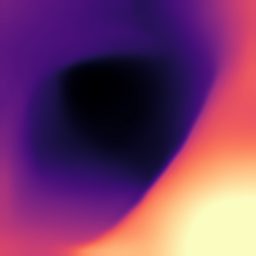} &
			\includegraphics[width=0.2\linewidth]{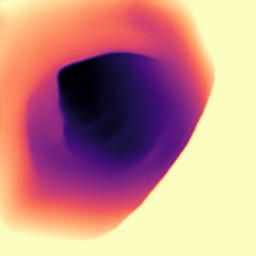} & \includegraphics[width=0.2\linewidth]{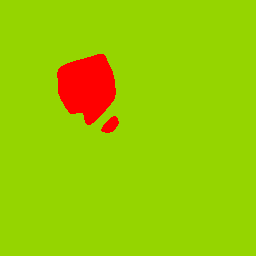} \\
			
			{\rotatebox{90}{\hspace{2mm}\shortstack{Endo \\ Syn\\ \cite{azagra2022endomapper}}}} &
			\includegraphics[width=0.2\linewidth]{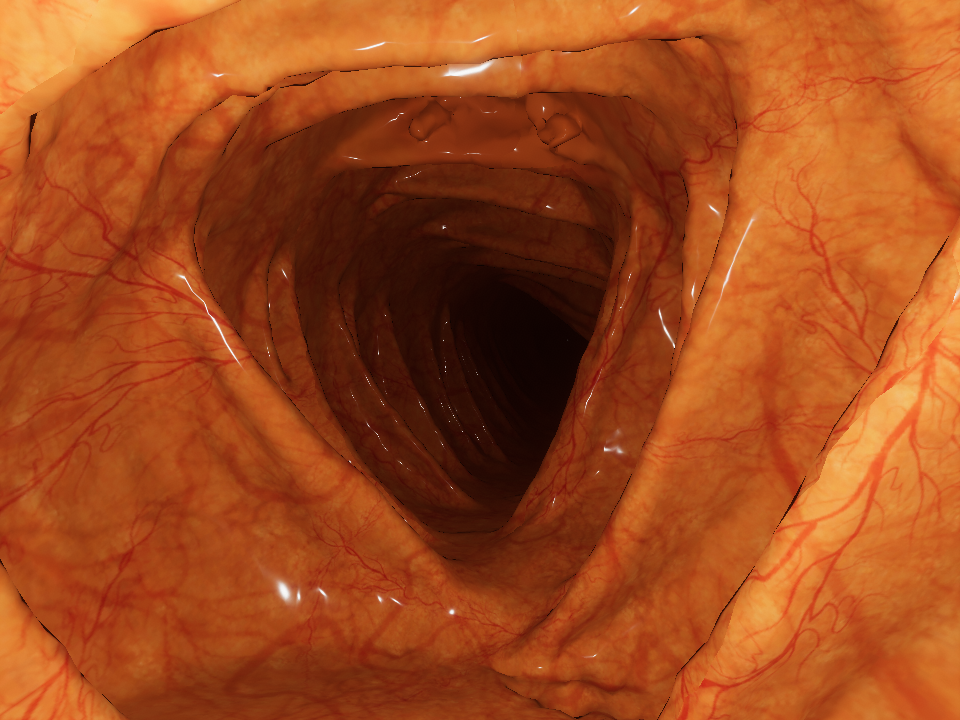} & \includegraphics[width=0.2\linewidth]{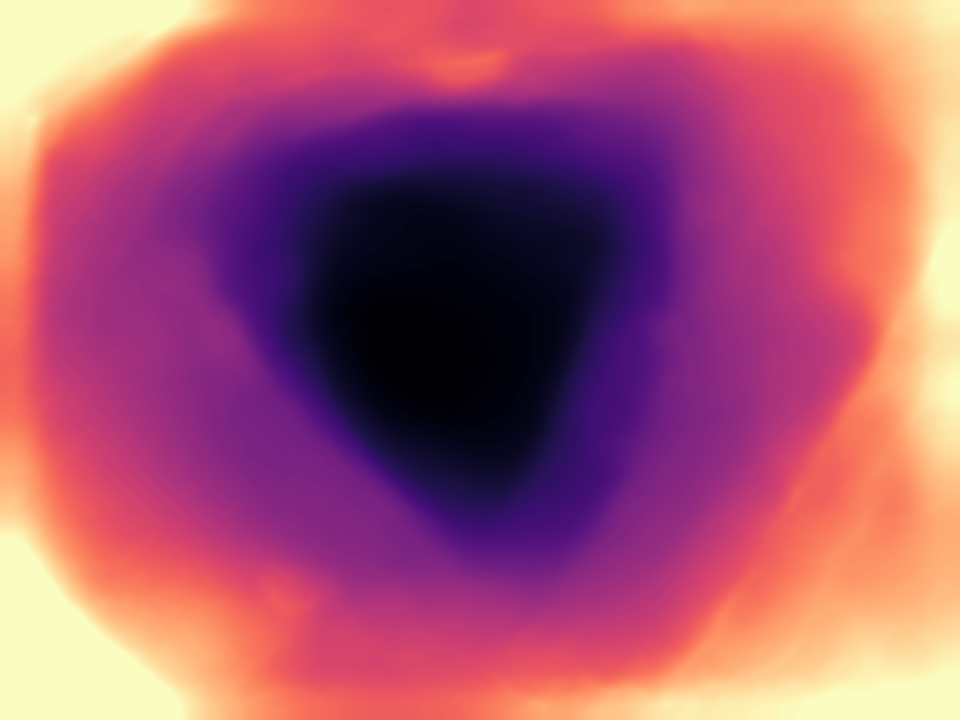} &
			\includegraphics[width=0.2\linewidth]{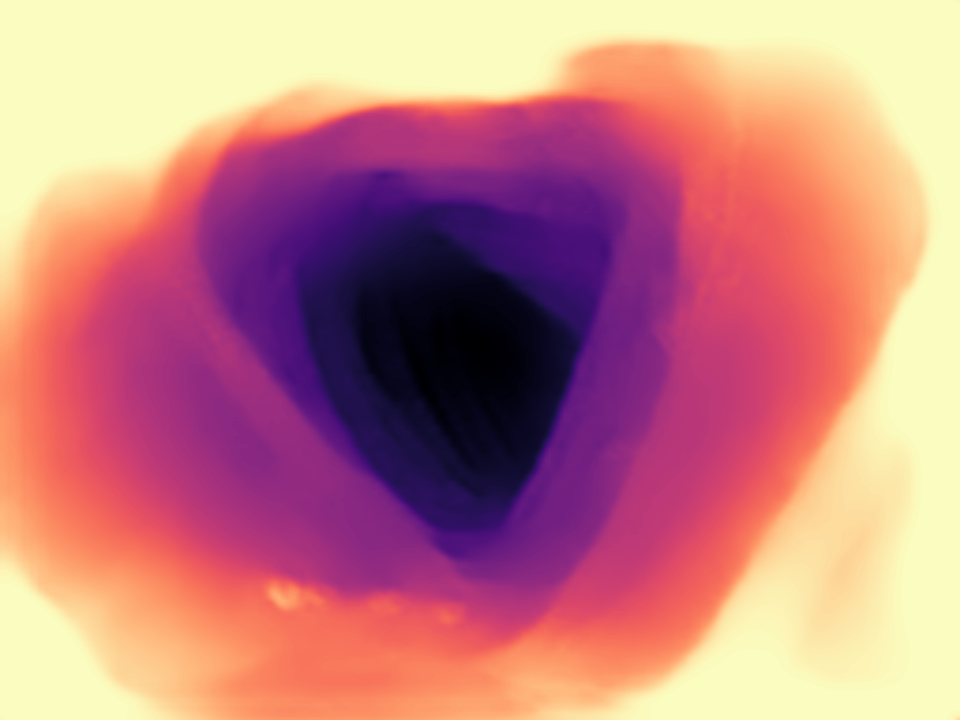} & \includegraphics[width=0.2\linewidth]{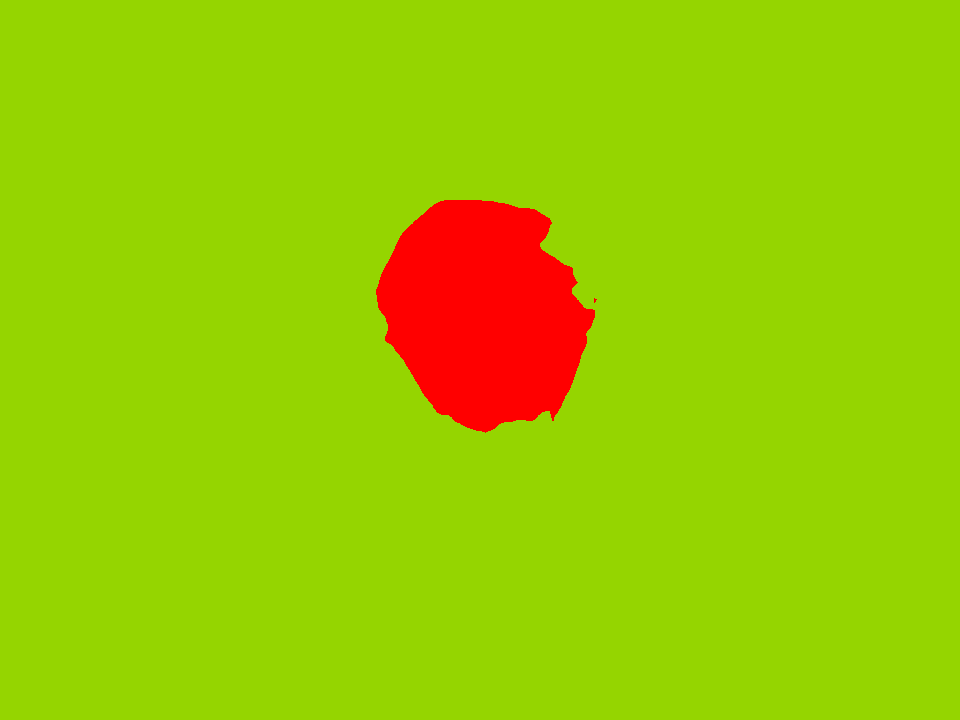} \\
			
			{\rotatebox{90}{\hspace{0.5mm}\shortstack{CTM \\dataset}}} &
			\includegraphics[width=0.2\linewidth]{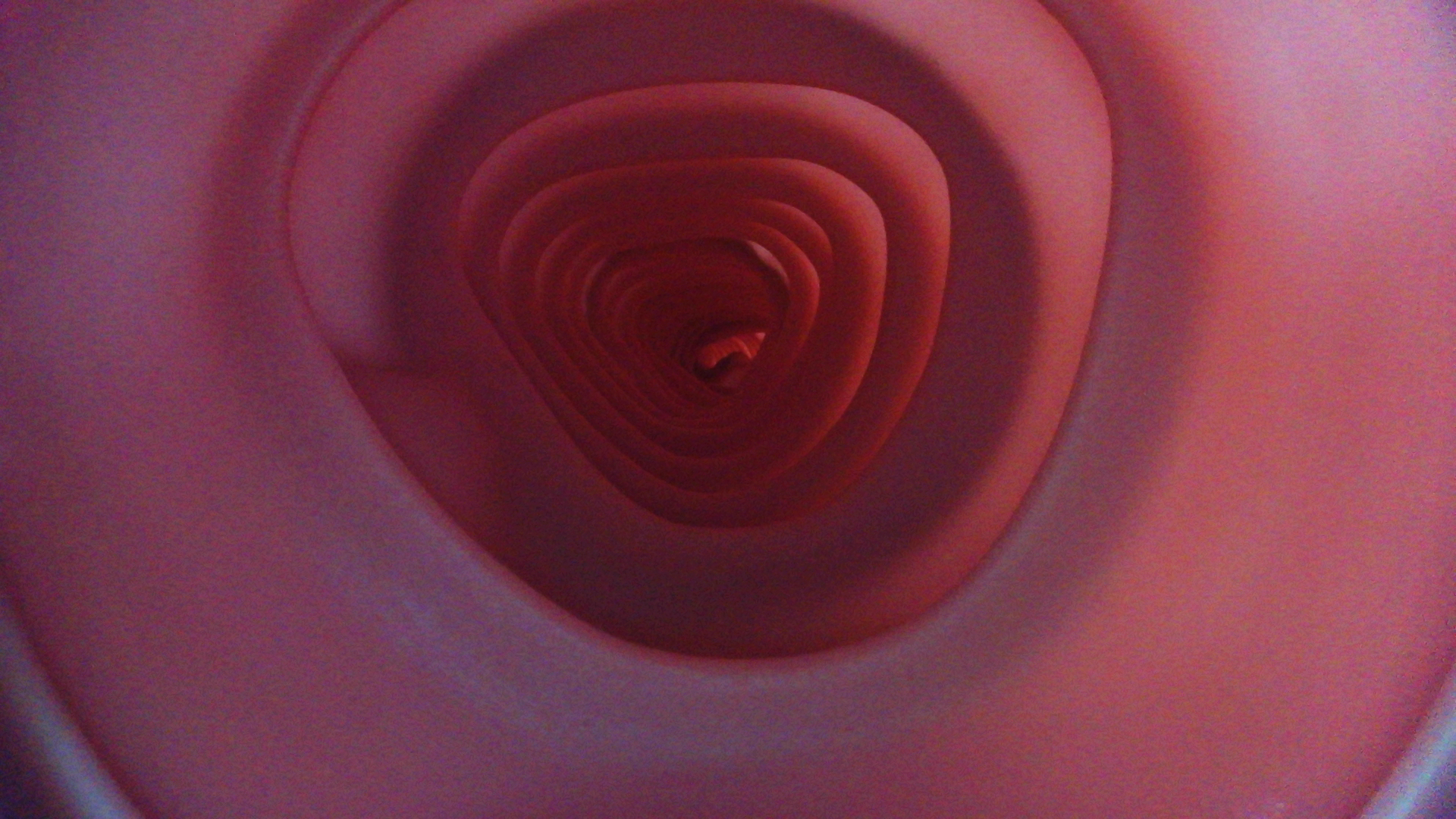} & \includegraphics[width=0.2\linewidth]{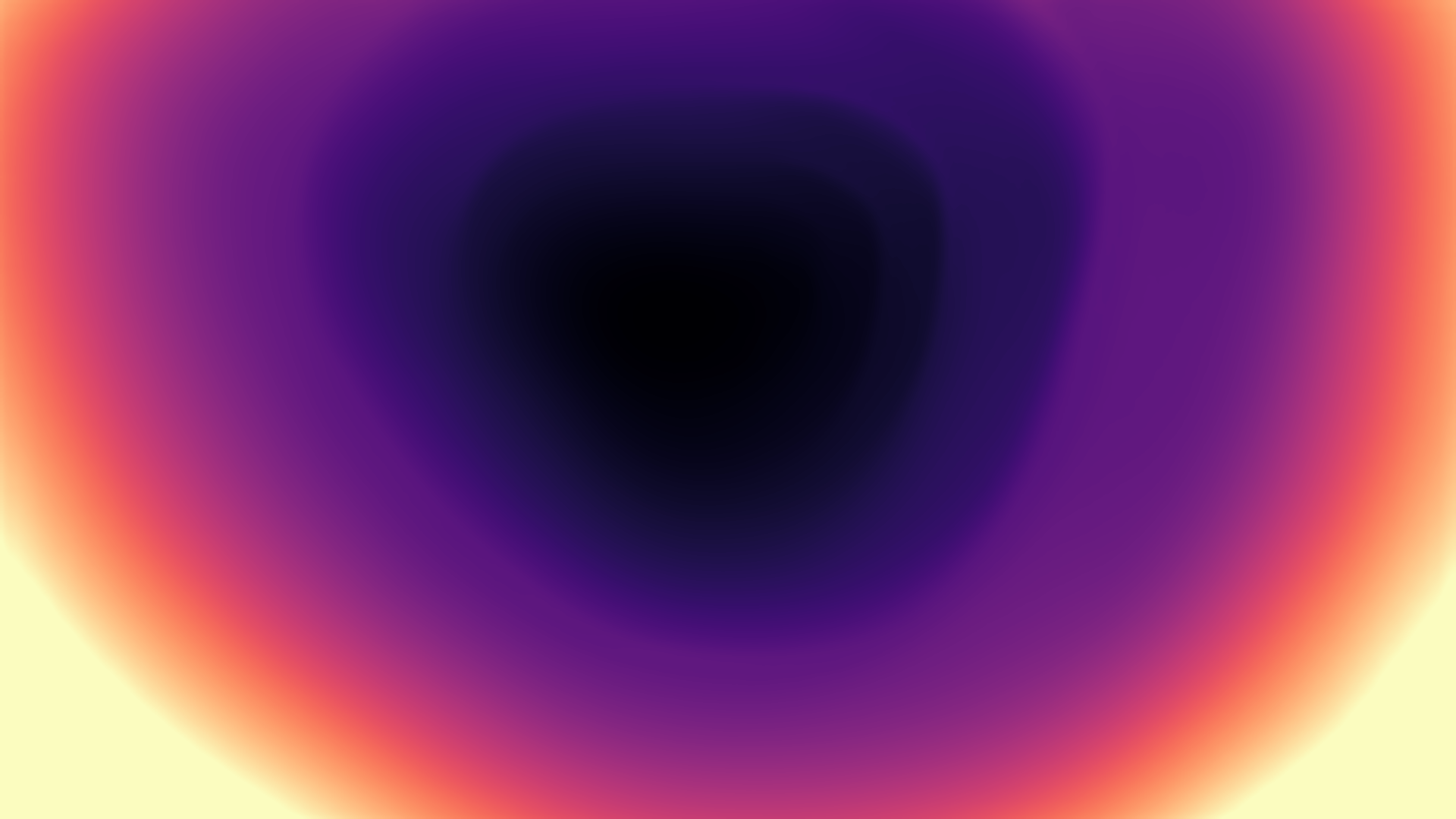} &
			\includegraphics[width=0.2\linewidth]{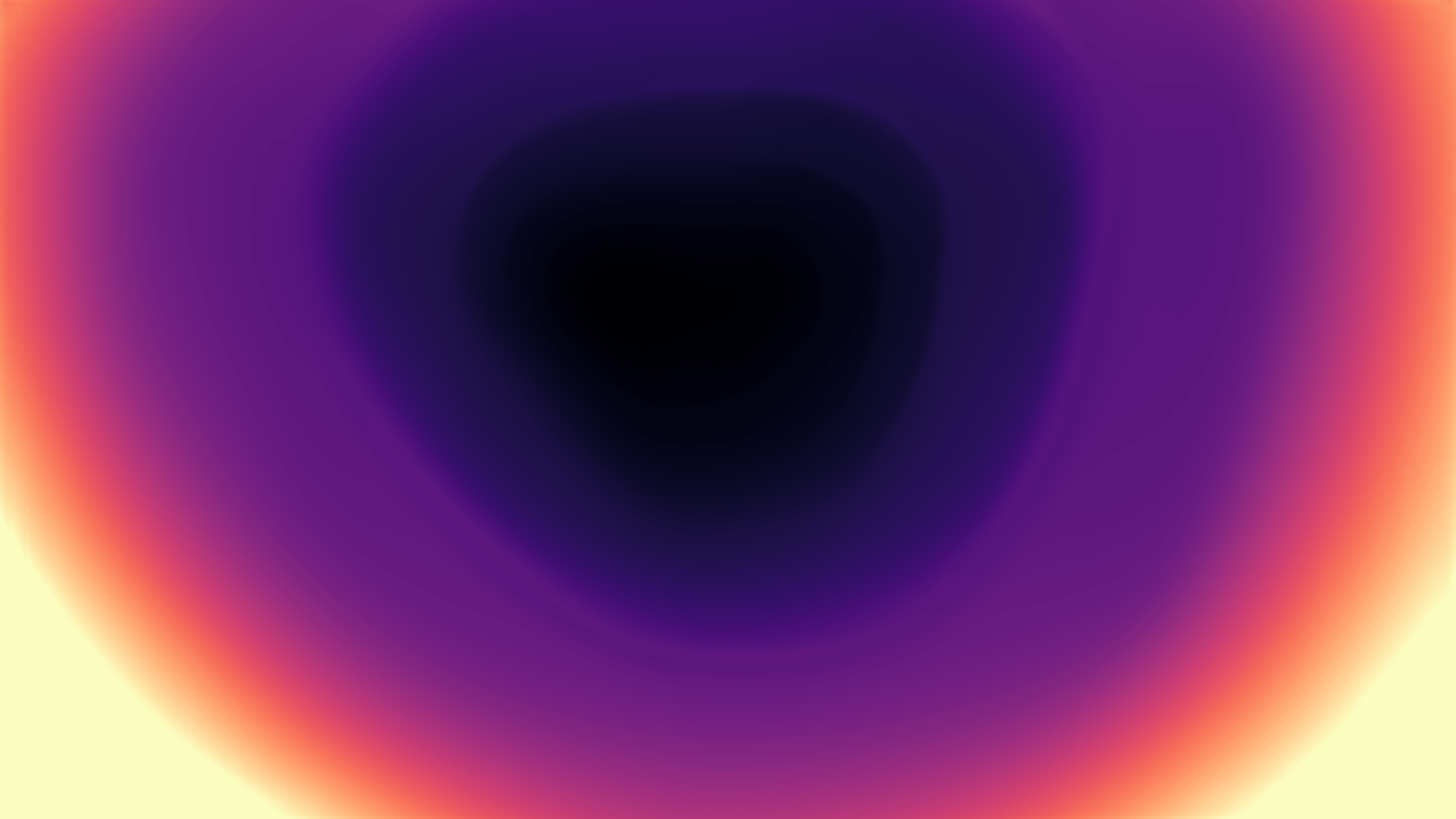} & \includegraphics[width=0.2\linewidth]{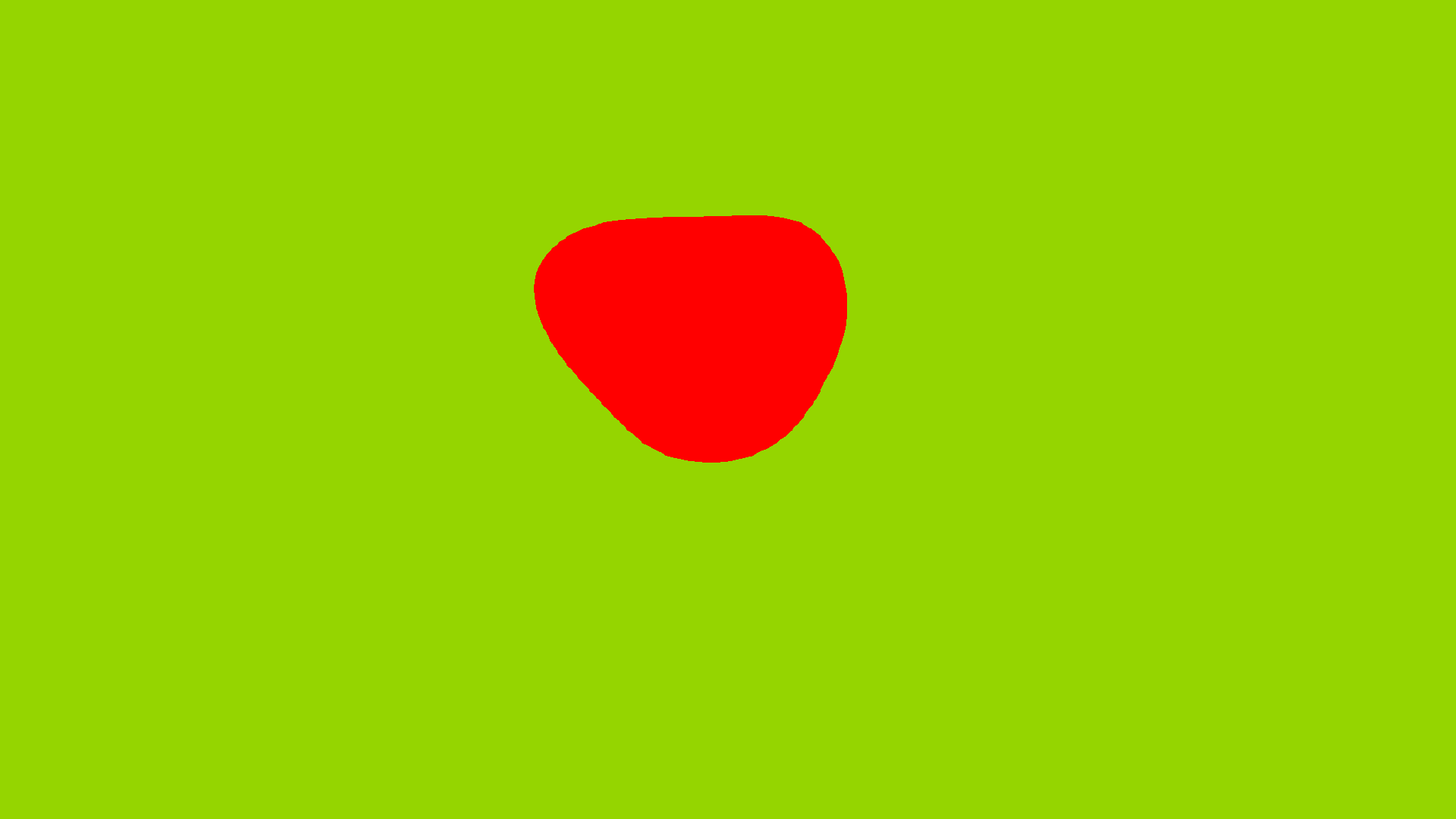} \\

			{\rotatebox{90}{\hspace{2mm}\shortstack{LDPoly \\ Real\\ \cite{ma2021ldpolypvideo}}}} &
			\includegraphics[width=0.2\linewidth]{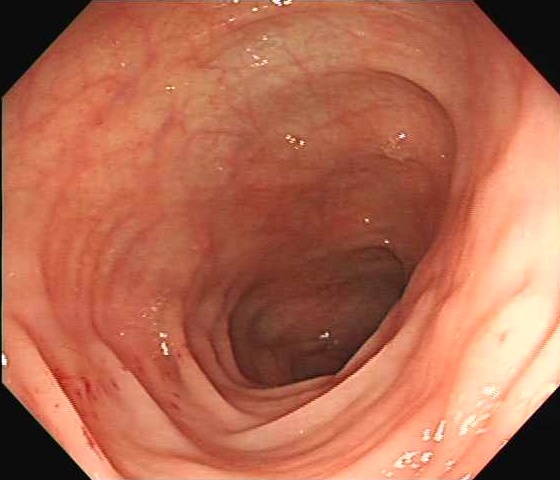} & \includegraphics[width=0.2\linewidth]{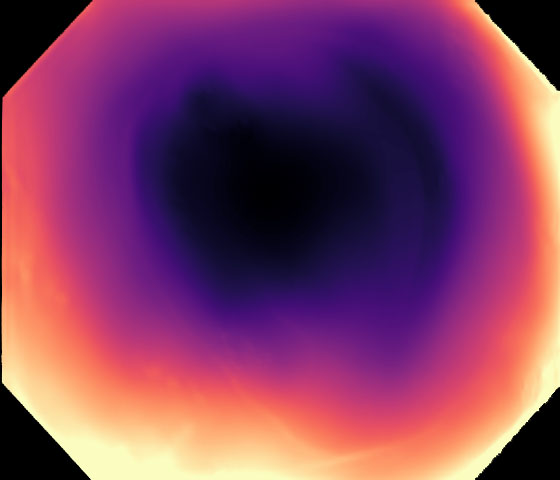} &
			\includegraphics[width=0.2\linewidth]{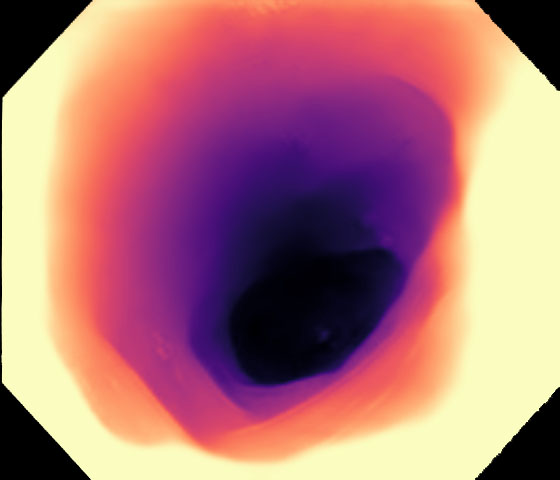} & \includegraphics[width=0.2\linewidth]{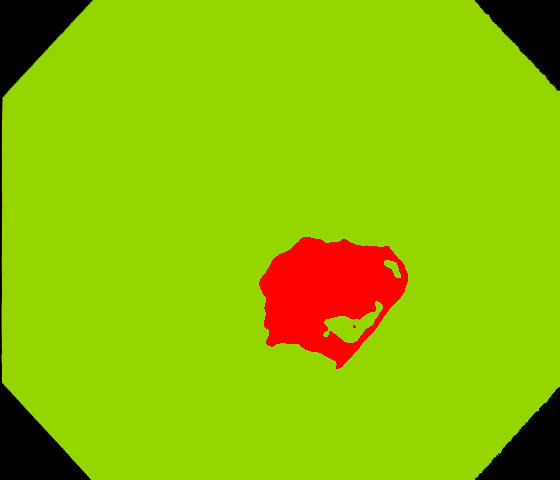} \\
			
			{\rotatebox{90}{\hspace{2.5mm}\shortstack{Endo\\ Real\\ \cite{azagra2022endomapper}}}} &
			\includegraphics[width=0.2\linewidth]{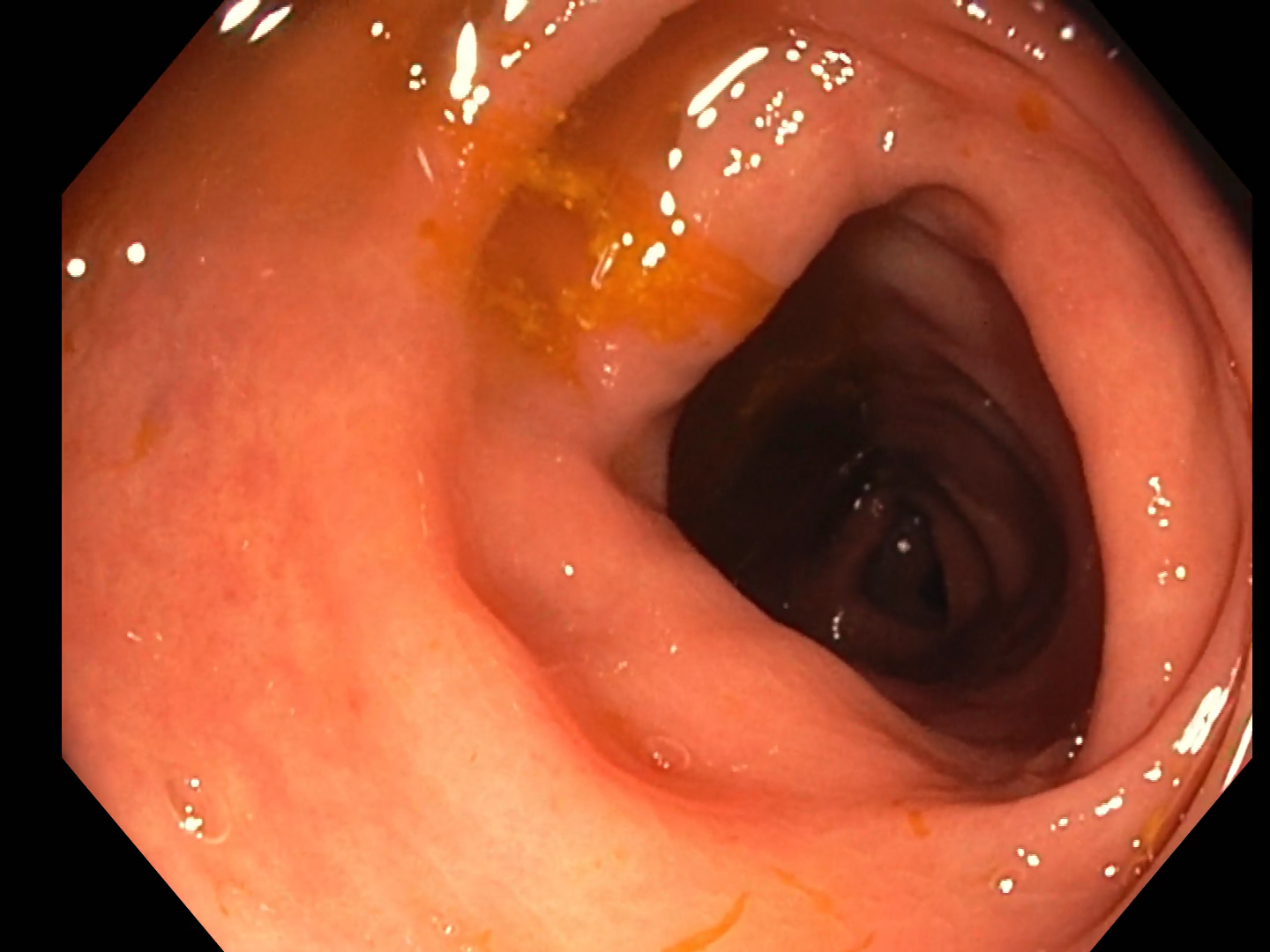} & \includegraphics[width=0.2\linewidth]{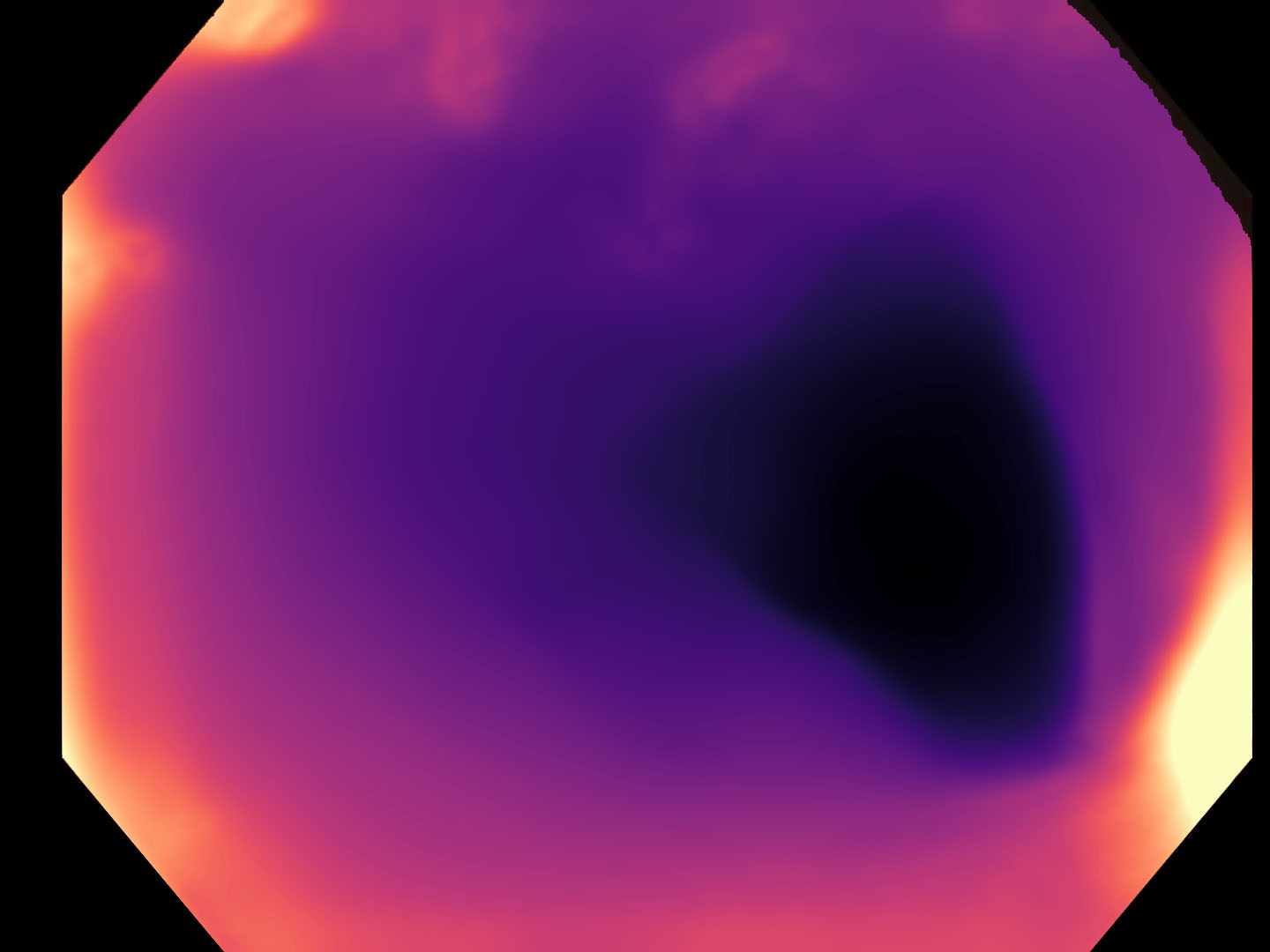} &
			\includegraphics[width=0.2\linewidth]{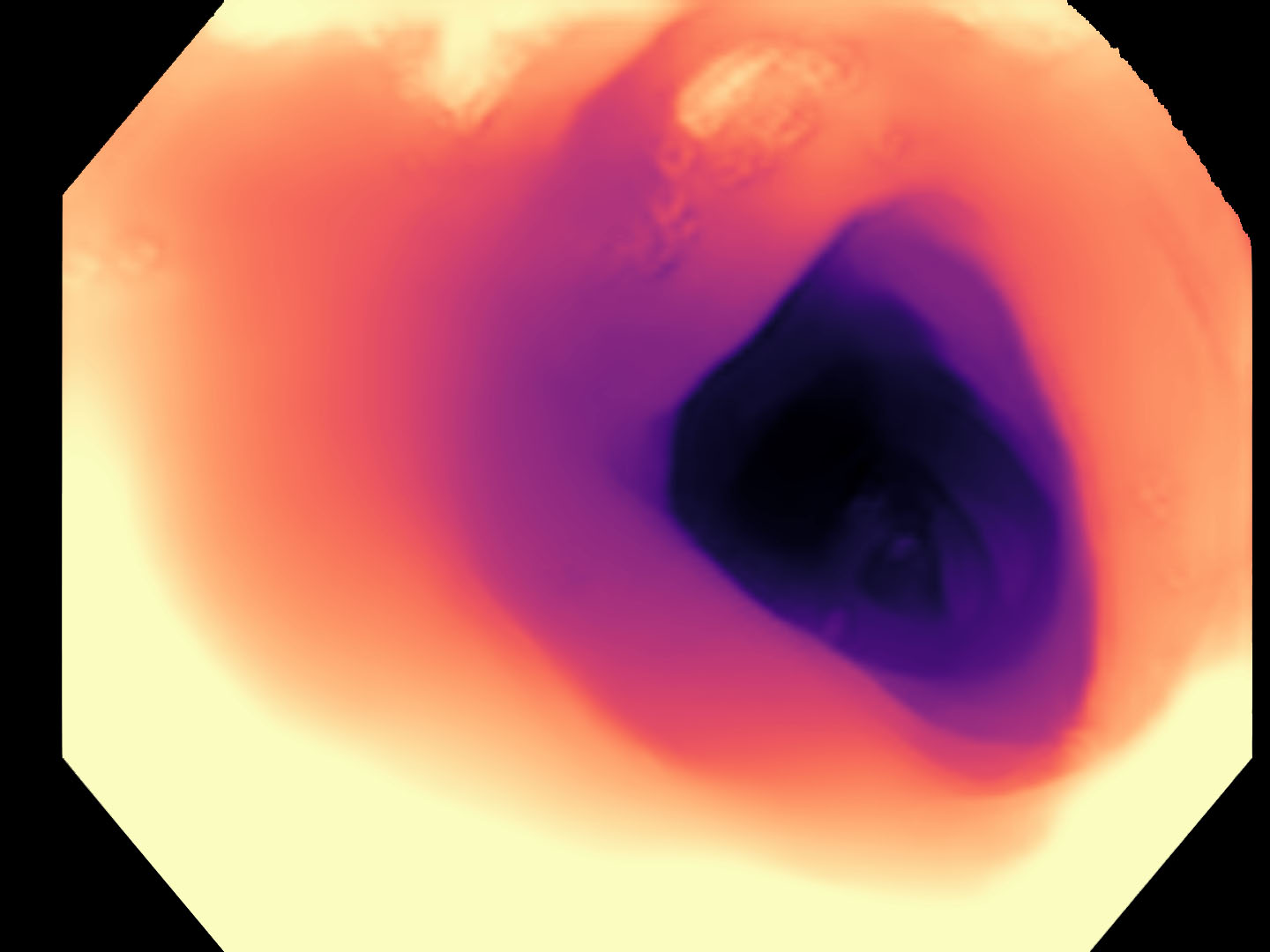} & \includegraphics[width=0.2\linewidth]{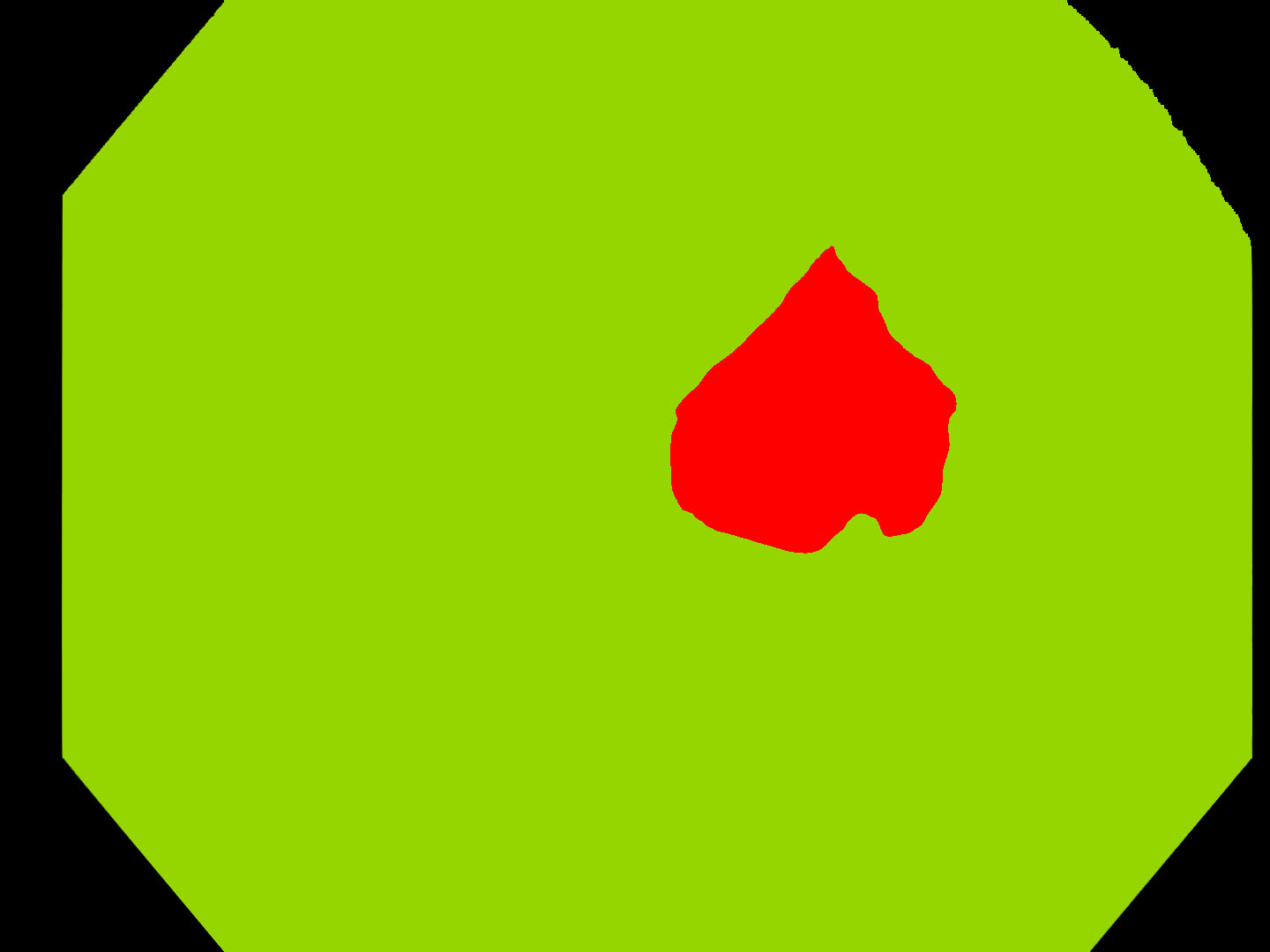} \\
			
		\end{tabular}
	}
	\caption{Qualitative analysis on synthetic datasets like UCL 2019 synthetic and Endomapper synthetic dataset, Colonoscopy Training Model (CTM) dataset, and real colonoscopy images like LDPolyVideo and Endomapper dataset (top to bottom). Baseline depth denotes depth estimated by depth network alone and SoftEnNet depth denotes depth estimated by SoftEnNet. SoftEnNet lumen shows the lumen segmentation (red-lumen and green-colon wall).}
	\label{fig:qual_1}
\end{figure}

\subsection{Quantitative study}

The unique structural restriction of real colon precludes the collection of ground truth depth with a true depth sensor. Thus, quantitative depth evaluation can only be conducted on synthetic data such as UCL 2019, even though the current synthetic data generation is not realistic enough to mimic real colon surface properties. The SoftEnNet depth estimation is assessed on UCL 2019 test dataset with evaluation metrics in \cite{eigen2014depth} and compare with other state-of-the-art methods in Table~\ref{tab:quant}. UCL synthetic data contains some texture, such as blood veins and haustral folds. The depth evaluation demonstrates the benefit of the validity mask, which ignores pixels that do not contribute to the photometric loss. It is also found that clipping off high photometric loss improves model performance. 

\subsection{Qualitative analysis}

It is challenging to evaluate the quantitative depth of real colonoscopy data such as Endomapper \cite{azagra2022endomapper}, LDPolypVideo \cite{ma2021ldpolypvideo}, or RVGAN \cite{itoh2021unsupervised} dataset. Many real-world characteristics of the colon, such as surface reflectance and light diffusion of different tissue types, are taxing to mimic in a simulator. Furthermore, a real colon is not rigid which complicates view synthesis even further. The depth network was found to leave holes in the depth maps when trained on a low-textured CTM with only image reconstruction loss. Low-textured regions are common in the real colon wall; how these regions are handled is critical for data-driven methods that rely on view synthesis. The validity mask described in Section \ref{photometric} effectively removes specular lighting, lens artifacts, and zero optical flow regions. The depth consistency loss described in Section \ref{consist} aids the network in learning geometric consistency from frame to frame and preventing flickering in depth maps. The depth network trained on CTM learns to predict depth directly from highly distorted images using the image synthesizer \ref{synthesis} with high FOV projection models like Double Sphere \cite{usenko2018double}. The qualitative analysis of the proposed SoftEnNet shown in Fig. \ref{fig:qual_1} clearly shows the symbiotic relation brings out more details in the depth estimation, especially on the haustral folds. Fig. \ref{fig:qual_2} shows a qualitative comparison of the proposed model depth prediction with state-of-the-art methods on real colonoscopy dataset \cite{itoh2021unsupervised}. The proposed model's robustness proves superior to that of RVGAN \cite{itoh2021unsupervised}, XDCycleGAN \cite{mathew2020augmenting}, CycleGAN \cite{zhu2017unpaired}, ExPix2Pix \cite{rau2019implicit}, Monodepth2 \cite{godard2019digging} and SfMLearner \cite{zhou2017unsupervised} with difficult real colonoscopy images.

\subsection{Ablation study}

Table~\ref{tab:task_wt} demonstrates the impact of learned task weights considering homoscedastic uncertainty against hand-crafted task weights. Table~\ref{tab:pac_alpha} shows the advantage of evolving PAC-$\alpha$ from a smaller initial value during training instead of a fixed alpha. Lower PAC-$\alpha$ indicates less information passed between sub-networks, but  PAC-$\alpha=1$ in the initial training stage makes the depth and lumen segmentation accuracy decrease. 

\section{CONCLUSION}

Understanding 3D and lumen tracking is essential for autonomous colonoscopy endorobot locomotion and surgical tasks. In this work, we developed a novel symbiotic network, SoftEnNet, that simultaneously predicts depth and lumen segmentation. The depth and lumen sub-network in SoftEnNet leverages the target task with guidance from the other task. We also introduced a novel validity mask that removes pixels affected by specular lighting, low texture surface, and violate Lambertian surface. 
This work can be used to implement autonomous tasks, which require a real-time closed-loop controller between the vision system and the device. The vision system proposed in this study can provide the direction of the locomotion to autonomous locomotion algorithms, i.e., the center of the lumen. 
Future planned works concerns the integration of the current vision system with the locomotion \cite{manfredi2019soft} and low-level controller \cite{manfredi2019wireless}. This includes testing the stability of the output against real camera movements, as these affect illumination (carried by the robot), hence image appearance, possibly reducing quality of the video as well as of the estimated depth maps and lumen location.

\bibliographystyle{IEEEtran}
\bibliography{egbib} 

\begin{thebibliography}{10}
\providecommand{\url}[1]{#1}
\csname url@rmstyle\endcsname
\providecommand{\newblock}{\relax}
\providecommand{\bibinfo}[2]{#2}
\providecommand\BIBentrySTDinterwordspacing{\spaceskip=0pt\relax}
\providecommand\BIBentryALTinterwordstretchfactor{4}
\providecommand\BIBentryALTinterwordspacing{\spaceskip=\fontdimen2\font plus
\BIBentryALTinterwordstretchfactor\fontdimen3\font minus
  \fontdimen4\font\relax}
\providecommand\BIBforeignlanguage[2]{{%
\expandafter\ifx\csname l@#1\endcsname\relax
\typeout{** WARNING: IEEEtran.bst: No hyphenation pattern has been}%
\typeout{** loaded for the language `#1'. Using the pattern for}%
\typeout{** the default language instead.}%
\else
\language=\csname l@#1\endcsname
\fi
#2}}

\bibitem{xi2021global}
Y.~Xi and P.~Xu, ``Global colorectal cancer burden in 2020 and projections to
  2040,'' \emph{Translational Oncology}, vol.~14, no.~10, p. 101174, 2021.

\bibitem{sung2021global}
H.~Sung, J.~Ferlay, R.~L. Siegel, M.~Laversanne, I.~Soerjomataram, A.~Jemal,
  and F.~Bray, ``Global cancer statistics 2020: Globocan estimates of incidence
  and mortality worldwide for 36 cancers in 185 countries,'' \emph{CA: a cancer
  journal for clinicians}, vol.~71, no.~3, pp. 209--249, 2021.

\bibitem{baxter2009association}
N.~N. Baxter, M.~A. Goldwasser, L.~F. Paszat, R.~Saskin, D.~R. Urbach, and
  L.~Rabeneck, ``Association of colonoscopy and death from colorectal cancer,''
  \emph{Annals of internal medicine}, vol. 150, no.~1, pp. 1--8, 2009.

\bibitem{manfredi2021endorobots}
L.~Manfredi, ``Endorobots for colonoscopy: Design challenges and available
  technologies,'' \emph{Frontiers in Robotics and AI}, p. 209, 2021.

\bibitem{manfredi2019soft}
L.~Manfredi, E.~Capoccia, G.~Ciuti, and A.~Cuschieri, ``A soft pneumatic
  inchworm double balloon (spid) for colonoscopy,'' \emph{Scientific reports},
  vol.~9, no.~1, pp. 1--9, 2019.

\bibitem{hong20143d}
D.~Hong, W.~Tavanapong, J.~Wong, J.~Oh, and P.~C. De~Groen, ``3d reconstruction
  of virtual colon structures from colonoscopy images,'' \emph{Computerized
  Medical Imaging and Graphics}, vol.~38, no.~1, pp. 22--33, 2014.

\bibitem{zhao2016endoscopogram}
Q.~Zhao, T.~Price, S.~Pizer, M.~Niethammer, R.~Alterovitz, and J.~Rosenman,
  ``The endoscopogram: A 3d model reconstructed from endoscopic video frames,''
  in \emph{International conference on medical image computing and
  computer-assisted intervention}.\hskip 1em plus 0.5em minus 0.4em\relax
  Springer, 2016, pp. 439--447.

\bibitem{widya2020stomach}
A.~R. Widya, Y.~Monno, M.~Okutomi, S.~Suzuki, T.~Gotoda, and K.~Miki, ``Stomach
  3d reconstruction based on virtual chromoendoscopic image generation,'' in
  \emph{2020 42nd Annual International Conference of the IEEE Engineering in
  Medicine \& Biology Society (EMBC)}.\hskip 1em plus 0.5em minus 0.4em\relax
  IEEE, 2020, pp. 1848--1852.

\bibitem{widya2019whole}
------, ``Whole stomach 3d reconstruction and frame localization from monocular
  endoscope video,'' \emph{IEEE Journal of Translational Engineering in Health
  and Medicine}, vol.~7, pp. 1--10, 2019.

\bibitem{widya2021self}
------, ``Self-supervised monocular depth estimation in gastroendoscopy using
  gan-augmented images,'' in \emph{Medical Imaging 2021: Image Processing},
  vol. 11596.\hskip 1em plus 0.5em minus 0.4em\relax SPIE, 2021, pp. 319--328.

\bibitem{ma2019real}
R.~Ma, R.~Wang, S.~Pizer, J.~Rosenman, S.~K. McGill, and J.-M. Frahm,
  ``Real-time 3d reconstruction of colonoscopic surfaces for determining
  missing regions,'' in \emph{International Conference on Medical Image
  Computing and Computer-Assisted Intervention}.\hskip 1em plus 0.5em minus
  0.4em\relax Springer, 2019, pp. 573--582.

\bibitem{freedman2020detecting}
D.~Freedman, Y.~Blau, L.~Katzir, A.~Aides, I.~Shimshoni, D.~Veikherman,
  T.~Golany, A.~Gordon, G.~Corrado, Y.~Matias, \emph{et~al.}, ``Detecting
  deficient coverage in colonoscopies,'' \emph{IEEE Transactions on Medical
  Imaging}, vol.~39, no.~11, pp. 3451--3462, 2020.

\bibitem{mahmood2018deep}
F.~Mahmood and N.~J. Durr, ``Deep learning and conditional random fields-based
  depth estimation and topographical reconstruction from conventional
  endoscopy,'' \emph{Medical image analysis}, vol.~48, pp. 230--243, 2018.

\bibitem{visentini2017deep}
M.~Visentini-Scarzanella, T.~Sugiura, T.~Kaneko, and S.~Koto, ``Deep monocular
  3d reconstruction for assisted navigation in bronchoscopy,''
  \emph{International journal of computer assisted radiology and surgery},
  vol.~12, no.~7, pp. 1089--1099, 2017.

\bibitem{bernal2012towards}
J.~Bernal, J.~S{\'a}nchez, and F.~Vilarino, ``Towards automatic polyp detection
  with a polyp appearance model,'' \emph{Pattern Recognition}, vol.~45, no.~9,
  pp. 3166--3182, 2012.

\bibitem{ganz2012automatic}
M.~Ganz, X.~Yang, and G.~Slabaugh, ``Automatic segmentation of polyps in
  colonoscopic narrow-band imaging data,'' \emph{IEEE Transactions on
  Biomedical Engineering}, vol.~59, no.~8, pp. 2144--2151, 2012.

\bibitem{tajbakhsh2015automated}
N.~Tajbakhsh, S.~R. Gurudu, and J.~Liang, ``Automated polyp detection in
  colonoscopy videos using shape and context information,'' \emph{IEEE
  transactions on medical imaging}, vol.~35, no.~2, pp. 630--644, 2015.

\bibitem{wang2018development}
P.~Wang, X.~Xiao, J.~R. Glissen~Brown, T.~M. Berzin, M.~Tu, F.~Xiong, X.~Hu,
  P.~Liu, Y.~Song, D.~Zhang, \emph{et~al.}, ``Development and validation of a
  deep-learning algorithm for the detection of polyps during colonoscopy,''
  \emph{Nature biomedical engineering}, vol.~2, no.~10, pp. 741--748, 2018.

\bibitem{guo2019giana}
Y.~B. Guo and B.~Matuszewski, ``Giana polyp segmentation with fully
  convolutional dilation neural networks,'' in \emph{Proceedings of the 14th
  International Joint Conference on Computer Vision, Imaging and Computer
  Graphics Theory and Applications}.\hskip 1em plus 0.5em minus 0.4em\relax
  SCITEPRESS-Science and Technology Publications, 2019, pp. 632--641.

\bibitem{jha2019resunet++}
D.~Jha, P.~H. Smedsrud, M.~A. Riegler, D.~Johansen, T.~De~Lange, P.~Halvorsen,
  and H.~D. Johansen, ``Resunet++: An advanced architecture for medical image
  segmentation,'' in \emph{2019 IEEE International Symposium on Multimedia
  (ISM)}.\hskip 1em plus 0.5em minus 0.4em\relax IEEE, 2019, pp. 225--2255.

\bibitem{wang2022boundary}
R.~Wang, S.~Chen, C.~Ji, J.~Fan, and Y.~Li, ``Boundary-aware context neural
  network for medical image segmentation,'' \emph{Medical Image Analysis},
  vol.~78, p. 102395, 2022.

\bibitem{jha2020doubleu}
D.~Jha, M.~A. Riegler, D.~Johansen, P.~Halvorsen, and H.~D. Johansen,
  ``Doubleu-net: A deep convolutional neural network for medical image
  segmentation,'' in \emph{2020 IEEE 33rd International symposium on
  computer-based medical systems (CBMS)}.\hskip 1em plus 0.5em minus
  0.4em\relax IEEE, 2020, pp. 558--564.

\bibitem{huang2021hardnet}
C.-H. Huang, H.-Y. Wu, and Y.-L. Lin, ``Hardnet-mseg: A simple encoder-decoder
  polyp segmentation neural network that achieves over 0.9 mean dice and 86
  fps,'' \emph{arXiv preprint arXiv:2101.07172}, 2021.

\bibitem{chao2019hardnet}
P.~Chao, C.-Y. Kao, Y.-S. Ruan, C.-H. Huang, and Y.-L. Lin, ``Hardnet: A low
  memory traffic network,'' in \emph{Proceedings of the IEEE/CVF international
  conference on computer vision}, 2019, pp. 3552--3561.

\bibitem{tomar2022fanet}
N.~K. Tomar, D.~Jha, M.~A. Riegler, H.~D. Johansen, D.~Johansen, J.~Rittscher,
  P.~Halvorsen, and S.~Ali, ``Fanet: A feedback attention network for improved
  biomedical image segmentation,'' \emph{IEEE Transactions on Neural Networks
  and Learning Systems}, 2022.

\bibitem{garg2016unsupervised}
R.~Garg, V.~K. Bg, G.~Carneiro, and I.~Reid, ``Unsupervised cnn for single view
  depth estimation: Geometry to the rescue,'' in \emph{European conference on
  computer vision}.\hskip 1em plus 0.5em minus 0.4em\relax Springer, 2016, pp.
  740--756.

\bibitem{godard2017unsupervised}
C.~Godard, O.~Mac~Aodha, and G.~J. Brostow, ``Unsupervised monocular depth
  estimation with left-right consistency,'' in \emph{Proceedings of the IEEE
  conference on computer vision and pattern recognition}, 2017, pp. 270--279.

\bibitem{jaderberg2015spatial}
M.~Jaderberg, K.~Simonyan, A.~Zisserman, \emph{et~al.}, ``Spatial transformer
  networks,'' \emph{Advances in neural information processing systems},
  vol.~28, 2015.

\bibitem{zhou2017unsupervised}
T.~Zhou, M.~Brown, N.~Snavely, and D.~G. Lowe, ``Unsupervised learning of depth
  and ego-motion from video,'' in \emph{Proceedings of the IEEE conference on
  computer vision and pattern recognition}, 2017, pp. 1851--1858.

\bibitem{mahjourian2018unsupervised}
R.~Mahjourian, M.~Wicke, and A.~Angelova, ``Unsupervised learning of depth and
  ego-motion from monocular video using 3d geometric constraints,'' in
  \emph{Proceedings of the IEEE conference on computer vision and pattern
  recognition}, 2018, pp. 5667--5675.

\bibitem{yin2018geonet}
Z.~Yin and J.~Shi, ``Geonet: Unsupervised learning of dense depth, optical flow
  and camera pose,'' in \emph{Proceedings of the IEEE conference on computer
  vision and pattern recognition}, 2018, pp. 1983--1992.

\bibitem{wang2004image}
Z.~Wang, A.~C. Bovik, H.~R. Sheikh, and E.~P. Simoncelli, ``Image quality
  assessment: from error visibility to structural similarity,'' \emph{IEEE
  transactions on image processing}, vol.~13, no.~4, pp. 600--612, 2004.

\bibitem{zhou2018unsupervised}
L.~Zhou, J.~Ye, M.~Abello, S.~Wang, and M.~Kaess, ``Unsupervised learning of
  monocular depth estimation with bundle adjustment, super-resolution and clip
  loss,'' \emph{arXiv preprint arXiv:1812.03368}, 2018.

\bibitem{wang2018learning}
C.~Wang, J.~M. Buenaposada, R.~Zhu, and S.~Lucey, ``Learning depth from
  monocular videos using direct methods,'' in \emph{Proceedings of the IEEE
  conference on computer vision and pattern recognition}, 2018, pp. 2022--2030.

\bibitem{godard2019digging}
C.~Godard, O.~Mac~Aodha, M.~Firman, and G.~J. Brostow, ``Digging into
  self-supervised monocular depth estimation,'' in \emph{Proceedings of the
  IEEE/CVF International Conference on Computer Vision}, 2019, pp. 3828--3838.

\bibitem{lyu2021hr}
X.~Lyu, L.~Liu, M.~Wang, X.~Kong, L.~Liu, Y.~Liu, X.~Chen, and Y.~Yuan,
  ``Hr-depth: High resolution self-supervised monocular depth estimation,'' in
  \emph{Proceedings of the AAAI Conference on Artificial Intelligence},
  vol.~35, no.~3, 2021, pp. 2294--2301.

\bibitem{shao2021self}
S.~Shao, Z.~Pei, W.~Chen, B.~Zhang, X.~Wu, D.~Sun, and D.~Doermann,
  ``Self-supervised learning for monocular depth estimation on minimally
  invasive surgery scenes,'' in \emph{2021 IEEE International Conference on
  Robotics and Automation (ICRA)}.\hskip 1em plus 0.5em minus 0.4em\relax IEEE,
  2021, pp. 7159--7165.

\bibitem{shao2022self}
S.~Shao, Z.~Pei, W.~Chen, W.~Zhu, X.~Wu, D.~Sun, and B.~Zhang,
  ``Self-supervised monocular depth and ego-motion estimation in endoscopy:
  appearance flow to the rescue,'' \emph{Medical image analysis}, vol.~77, p.
  102338, 2022.

\bibitem{bian2021unsupervised}
J.-W. Bian, H.~Zhan, N.~Wang, Z.~Li, L.~Zhang, C.~Shen, M.-M. Cheng, and
  I.~Reid, ``Unsupervised scale-consistent depth learning from video,''
  \emph{International Journal of Computer Vision}, vol. 129, no.~9, pp.
  2548--2564, 2021.

\bibitem{kendall2018multi}
A.~Kendall, Y.~Gal, and R.~Cipolla, ``Multi-task learning using uncertainty to
  weigh losses for scene geometry and semantics,'' in \emph{Proceedings of the
  IEEE conference on computer vision and pattern recognition}, 2018, pp.
  7482--7491.

\bibitem{guizilini2020semanticallyguided}
V.~Guizilini, R.~Hou, J.~Li, R.~Ambrus, and A.~Gaidon, ``Semantically-guided
  representation learning for self-supervised monocular depth,'' in
  \emph{ICLR}, 2020.

\bibitem{su2019pixel}
H.~Su, V.~Jampani, D.~Sun, O.~Gallo, E.~Learned-Miller, and J.~Kautz,
  ``Pixel-adaptive convolutional neural networks,'' in \emph{Proceedings of the
  IEEE/CVF Conference on Computer Vision and Pattern Recognition}, 2019, pp.
  11\,166--11\,175.

\bibitem{agarap2018deep}
A.~F. Agarap, ``Deep learning using rectified linear units (relu),''
  \emph{arXiv preprint arXiv:1803.08375}, 2018.

\bibitem{he2016deep}
K.~He, X.~Zhang, S.~Ren, and J.~Sun, ``Deep residual learning for image
  recognition,'' in \emph{Proceedings of the IEEE conference on computer vision
  and pattern recognition}, 2016, pp. 770--778.

\bibitem{paszke2017automatic}
A.~Paszke, S.~Gross, S.~Chintala, G.~Chanan, E.~Yang, Z.~DeVito, Z.~Lin,
  A.~Desmaison, L.~Antiga, and A.~Lerer, ``Automatic differentiation in
  pytorch,'' in \emph{NIPS-W}, 2017.

\bibitem{kingma2014adam}
D.~P. Kingma and J.~Ba, ``Adam: A method for stochastic optimization,''
  \emph{arXiv preprint arXiv:1412.6980}, 2014.

\bibitem{rau2019implicit}
A.~Rau, P.~Edwards, O.~F. Ahmad, P.~Riordan, M.~Janatka, L.~B. Lovat, and
  D.~Stoyanov, ``Implicit domain adaptation with conditional generative
  adversarial networks for depth prediction in endoscopy,'' \emph{International
  journal of computer assisted radiology and surgery}, vol.~14, no.~7, pp.
  1167--1176, 2019.

\bibitem{itoh2021unsupervised}
H.~Itoh, M.~Oda, Y.~Mori, M.~Misawa, S.-E. Kudo, K.~Imai, S.~Ito, K.~Hotta,
  H.~Takabatake, M.~Mori, \emph{et~al.}, ``Unsupervised colonoscopic depth
  estimation by domain translations with a lambertian-reflection keeping
  auxiliary task,'' \emph{International Journal of Computer Assisted Radiology
  and Surgery}, vol.~16, no.~6, pp. 989--1001, 2021.

\bibitem{mathew2020augmenting}
S.~Mathew, S.~Nadeem, S.~Kumari, and A.~Kaufman, ``Augmenting colonoscopy using
  extended and directional cyclegan for lossy image translation,'' in
  \emph{Proceedings of the IEEE/CVF Conference on Computer Vision and Pattern
  Recognition}, 2020, pp. 4696--4705.

\bibitem{zhu2017unpaired}
J.-Y. Zhu, T.~Park, P.~Isola, and A.~A. Efros, ``Unpaired image-to-image
  translation using cycle-consistent adversarial networks,'' in
  \emph{Proceedings of the IEEE international conference on computer vision},
  2017, pp. 2223--2232.

\bibitem{azagra2022endomapper}
P.~Azagra, C.~Sostres, {\'A}.~Ferrandez, L.~Riazuelo, C.~Tomasini, O.~L.
  Barbed, J.~Morlana, D.~Recasens, V.~M. Batlle, J.~J.
  G{\'o}mez-Rodr{\'\i}guez, \emph{et~al.}, ``Endomapper dataset of complete
  calibrated endoscopy procedures,'' \emph{arXiv preprint arXiv:2204.14240},
  2022.

\bibitem{ma2021ldpolypvideo}
Y.~Ma, X.~Chen, K.~Cheng, Y.~Li, and B.~Sun, ``Ldpolypvideo benchmark: A
  large-scale colonoscopy video dataset of diverse polyps,'' in
  \emph{International Conference on Medical Image Computing and
  Computer-Assisted Intervention}.\hskip 1em plus 0.5em minus 0.4em\relax
  Springer, 2021, pp. 387--396.

\bibitem{eigen2014depth}
D.~Eigen, C.~Puhrsch, and R.~Fergus, ``Depth map prediction from a single image
  using a multi-scale deep network,'' \emph{Advances in neural information
  processing systems}, vol.~27, 2014.

\bibitem{usenko2018double}
V.~Usenko, N.~Demmel, and D.~Cremers, ``The double sphere camera model,'' in
  \emph{2018 International Conference on 3D Vision (3DV)}.\hskip 1em plus 0.5em
  minus 0.4em\relax IEEE, 2018, pp. 552--560.

\bibitem{manfredi2019wireless}
L.~Manfredi and A.~Cuschieri, ``A wireless compact control unit (wiccu) for
  untethered pneumatic soft robots,'' in \emph{2019 2nd IEEE International
  Conference on Soft Robotics (RoboSoft)}.\hskip 1em plus 0.5em minus
  0.4em\relax IEEE, 2019, pp. 31--36.

\end{thebibliography}

\end{document}